\documentclass[aps,reprint,showpacs,floatfix,longbibliography]{revtex4-1} 

\usepackage{amsmath}
\usepackage[english]{babel}
\usepackage{graphicx}
\usepackage{multirow}
\usepackage{subfig}

\begin{document}

\title{Asymmetries arising from the space-filling nature of vascular networks}

\author{David Hunt}
\author{Van M. Savage}
\affiliation{Department of Biomathematics, University of California at Los Angeles, Los Angeles, CA 90095, USA}

\date{\today}

\begin{abstract}
Cardiovascular networks span the body by branching across many generations of vessels.
The resulting structure delivers blood over long distances to supply all cells with oxygen via the relatively short-range process of diffusion at the capillary level.
The structural features of the network that accomplish this density and ubiquity of capillaries are often called \textit{space-filling}.
There are multiple strategies to fill a space, but some strategies do not lead to biologically adaptive structures by requiring too much construction material or space, delivering resources too slowly, or using too much power to move blood through the system.
We empirically measure the structure of real networks (18 humans and 1 mouse) and compare these observations with predictions of model networks that are space-filling and constrained by a few guiding biological principles.
We devise a numerical method that enables the investigation of space-filling strategies and determination of which biological principles influence network structure.
Optimization for only a single principle creates unrealistic networks that represent an extreme limit of the possible structures that could be observed in nature.
We first study these extreme limits for two competing principles, minimal total material and minimal path lengths.
We combine these two principles and enforce various thresholds for balance in the network hierarchy, which provides a novel approach that highlights the trade-offs faced by biological networks and yields predictions that better match our empirical data.
\end{abstract}



\maketitle

\section{Introduction}

The vital functions of the cardiovascular system are the distribution of oxygen and nutrient resources throughout the body, as well as the collection and filtration of waste by circulating blood.
Transfer of resources and waste occurs primarily at the capillary level via diffusion through nearby tissue.
Consequently, this smallest level of the network must reach all living cells in order to maintain them, filling the entire space of the body.
In models developed by Krogh for effective diffusion of oxygen \cite{Krogh1919}, cells cannot survive beyond a maximum distance from a capillary.
This defines a service volume of cells that is associated with each capillary, which has a typical size that has been observed to vary across species based on cellular metabolic rate \cite{West_1997, SavagePLoS}.
The constraint on maximum distance from capillaries necessitates that the final levels of the cardiovascular network are also space-filling throughout the body.
In this paper we investigate the relation between this space-filling property and basic optimization principles such as the minimization of costs from construction material and pumping power.
Specifically, we highlight how this relation influences the asymmetries in sizes and flow rates of sibling segments as measured in empirical data.

A central focus of our investigation of cardiovascular systems is the space-filling properties of networks, but these properties are also of great interest in many other contexts.
General space-covering hexagonal patterns appear in nature in the cell structure of beehives as well as in economic theories for market areas \cite{Puu05}.
Trees (the woody, perennial plants) have been studied for both how forests fill an area \cite{West09}, as well as how the vascular structure within an individual plant determines the hydraulics of resource delivery \cite{Savage10, Sperry12}.
Apollonian networks \cite{Andrade05} describe the space-filling packing of spheres of various sizes, which are similar in the cardiovascular system to considering the volumes of different subtrees of the network.
Efficiently filling space in two dimensions is important for information visualization \cite{Stasko00}.
In addition to these applications, Kuffner and LaValle study space-filling tree networks (i.e., networks that branch and have no closed loops) to determine a route from one location to another \cite{Kuffner11}, but without the biological constraints that we impose here.
For cardiovascular networks, this motion planning is analogous to how the vascular structure routes blood.
Efficient routing of blood through a hierarchy is central to models that investigate allometric scaling of metabolic rate with body mass \cite{West_1997, Banavar99, Banavar02, SavagePLoS, Dodds10, Huo12}, which build on metabolic scaling by Kleiber \cite{Kleiber32}.
Determining specific space-filling strategies will inform these models to better describe the cardiovascular system.

Developmental processes (i.e., growth) as well as evolutionary pressures, such as efficiency in material and energy use, shape the structure of cardiovascular networks.
Filling a volume or surface efficiently with the terminal nodes of a branching network is nontrivial, especially when the distribution system must reliably deliver blood at each stage of development.
The system must also be robust to changes in vessel lengths and the number of hierarchical levels that result from growth and obstructions from damage \cite{Corson10, Katifori10}.
We propound that these structural challenges lead to the asymmetric features in both the local relations of segments, as well as in the global paths from the heart to each service volume that we observe in empirical data.

At the local level, the ratio of lengths between parent and child segments may vary across the network, deviating from strict self-similarity.
Additionally, sibling segments may vary in length, which we call \textit{asymmetric branching}.
Within our numerical method, these features arise as a result of optimizing branch point positions relative to adjacent branch points across the network.
Variation in the length and number of levels in paths means that the network is also not symmetric or \textit{balanced} in these global properties.
By allowing these asymmetries and explicitly ensuring space-filling structure, we expand other models that are strictly balanced in network hierarchy and perfectly symmetrical in local quantities.

Asymmetries observed in real systems motivate our investigation of the space-filling properties and asymmetries in cardiovascular networks.
These observations show that such networks have a tendency for the lengths of sibling segments to be distributed less symmetrically than is the case for radii.
The empirical data in Sec. \ref{results_section} come from two types of sources.
Images collected through microtomography of mouse lung comprise one data set.
The mice were part of a study with different manipulations of matrix GLA protein (which causes the vasculature to be under- or over-branched \cite{Yao07}), but we focus on the wild-type specimen for our analysis.
The other data set, collected through MRI, excludes the lungs and instead focuses on the central vasculature in the human head and torso \cite{Johnson15}.
We utilize the custom software Angicart \cite{Johnson15} to analyze these two distinct vascular data sets.
Because of the noninvasive nature of these data acquisition and analysis techniques, future studies have the opportunity to track the development of cardiovascular systems as individual organisms grow and age, including repair after the system incurs damage (i.e., wound healing).
Such studies can elucidate the effects that patterns of growth and changes from damage have on the final, mature state of the network.

In this paper, we study the optimization principles that correspond to evolutionary pressures for efficiency in material cost and power loss.
Our focus is the influence of space-filling patterns on length asymmetry distributions without the explicit inclusion of radius information.
The list of candidate networks includes all possible hierarchical (topological) connections between the heart and all capillaries.
For each hierarchy and unique permutation of pairings between terminal vessels and service volumes (see Fig. \ref{tree_search}), we must determine the positions of the branch points.
The combination of the hierarchy, service-volume pairings and branch point positions defines the configuration of a candidate network.
For these reasons, we must search through many candidate configurations to determine the most efficient structure.
We quantify the fitness of each candidate network using individual segment lengths between branch points as well as full path lengths between each capillary and the heart.

To perform a reliable comparison between candidate configurations, it is crucial to determine branch point positions in a consistent way.
We determine these positions iteratively for the entire network in order to identify the global optimum. 
While the local process of choosing branch points that minimize total vessel lengths (or similar features) is relatively straightforward to iterate over the network, any single branch point and its relation to its neighbors relies indirectly on updates that are applied elsewhere.
This dependence emerges from the fact that each end of a vessel is connected to a branch point, which upon repositioning affects the lengths of all vessels that it joins.
The uniqueness of the \textit{Fermat point} --- the branch point position that minimizes the sum of the lengths of vessels at a single junction --- is already well established (for example, see \cite{Shen08}).
This allows us to carefully construct an algorithm (described in Sec. \ref{local_optimization_of_positions_subsection}) that reliably relaxes all branch point positions into the global optimum.

After determining the positions of branch points for a given hierarchy, we compare distinct configurations to find the optimum network.
The search through configurations is also a central problem in phylogenetics, where the goal is to construct phylogenies to identify similar groups of species and trace the development of genes through speciation.
Even in the case of genes that control biological traits, a loosely analogous space-filling phenomenon emerges in the form of species filling the niches in the environment.
With our specific goal of complete spatial covering of network tips, we develop strategies in Sec. \ref{configuration_section} for exploring the space of hierarchies that are similar to those used on phylogenetic trees.

The organization of the subsequent sections is as follows.
In Sec. \ref{model_section}, we describe the basic assumptions for our space-filling network model, including the details of the local optimization of branch point locations and the global paths through the network.
In Sec. \ref{quantification_section}, we introduce the specific quantitative network properties that we use to compare the fitness of candidate networks.
We introduce the properties of the space of tree hierarchies and our implemented exploration strategies in Sec. \ref{configuration_section}.
In Sec. \ref{results_section}, we detail the results from the several layers of optimization that we implement, and discuss the insights that they offer in Sec. \ref{discussion_section}.

\section{Construction of artificial vascular networks}
\label{model_section}

To better understand the connection between the local asymmetries of individual vessel lengths and the global constraint on space-filling capillaries, we optimize candidate networks that are embedded in two spatial dimensions ({2-$D$}) \textit{in silico} with respect to specific optimization principles.
We explore these optimized artificial networks and quantify their branching length asymmetries to compare with our empirical data.
Our model's simplification of the cardiovascular network focuses on the lengths of segments as defined by the straight line between adjacent branch points.
Reticulated structures occur within leaves to mitigate damage \cite{Corson10, Katifori10} and within animals as anastomoses (or pathologically as fistulas).
However, we focus on the vast majority of the cardiovascular system that distributes resources through a hierarchical, tree-like structure, in which no segments or subtrees rejoin downstream to form closed loops before reaching the capillaries.
This is sufficient for the focus of our investigation of the asymmetric, space-filling structure that distributes the resource-rich blood from the heart throughout the body.

The space-filling property of the cardiovascular network constrains the hierarchical structure of the network and the positions of branch points.
Here we describe our process for the construction of individual networks and the space of possible networks through the following steps: defining a distribution of space-filling service volumes in the space of the body, identifying all unique hierarchies and pairings between tips of hierarchies and distinct service volumes, and determining the positions of branch points for each hierarchy and pairing.

\subsection{Space-filling service volumes}

Because real systems do not organize or grow on a regular (symmetric, isotropic) grid, we position service volumes randomly within the space they fill.
Construction of service volumes begins by choosing a random point within the body volume that represents the location of a capillary.
We then randomly choose other points (capillary locations) so that none lie within a predefined constant distance from another capillary location.
After determining a set of capillary locations that span the {$2$-D} area, the entire body is partitioned into Voronoi cells fed by the closest capillary.
In this way, each capillary becomes associated with a specific service volume, and the sum of the service volumes fills the entire space (see Fig. \ref{simpleNetworks} or \ref{large_optimal_networks_circ}).

\subsection{Space of hierarchically distinct trees and pairings with service volumes}

Because multiple branching levels connect the service volumes to the heart, there are many possible hierarchical orderings of branching junctions across these different levels.
For example, there are two unique hierarchies when there are four service volumes: the top three configurations in Fig. \ref{tree_search} are the same perfectly balanced hierarchy, while the remaining trees have the same unbalanced hierarchy.
The distinguishing feature is the pairings of tips in the hierarchy to specific service volumes (1-4).

There are many distinct pairings of terminal tips with the set of fixed service volumes for each bifurcating tree.
For four service volumes, there are a manageable 15 unique bifurcating trees (shown in Fig. \ref{tree_search}).
Before determining branch point positions for small networks, we exhaustively search through all possible hierarchies and pairings with service volumes.
For large networks, the number of distinct hierarchies and pairings ($(2n - 3)!!$ for $n$ distinct service volumes) is prohibitively large, so we sample the space as described in Sec. \ref{larger_networks_subsection}.

We do not disqualify configurations if one vessel path crosses with another (these would likely separate in three dimensions), and there is no exchange of resources or interaction in blood flow at such locations.
Crossings are not observed for networks that minimize only total network length without a constraint on hierarchical balance, but they often occur for optimal configurations with a strong constraint on hierarchical balance.

\subsection{Optimization of branch point positions for a fixed hierarchy and pairing}
\label{local_optimization_of_positions_subsection}

\begin{figure*}[ht]
\vspace*{0.5cm}
\centering
\begin{tabular}[t]{c|c}
&
\parbox{6cm}{
	\centering{
		Global Optimization via Search through the Space of Hierarchies
	}
}
\\
\multirow{-2}{4cm}[1cm]{
	\subfloat[]{
		\includegraphics[width=4cm]{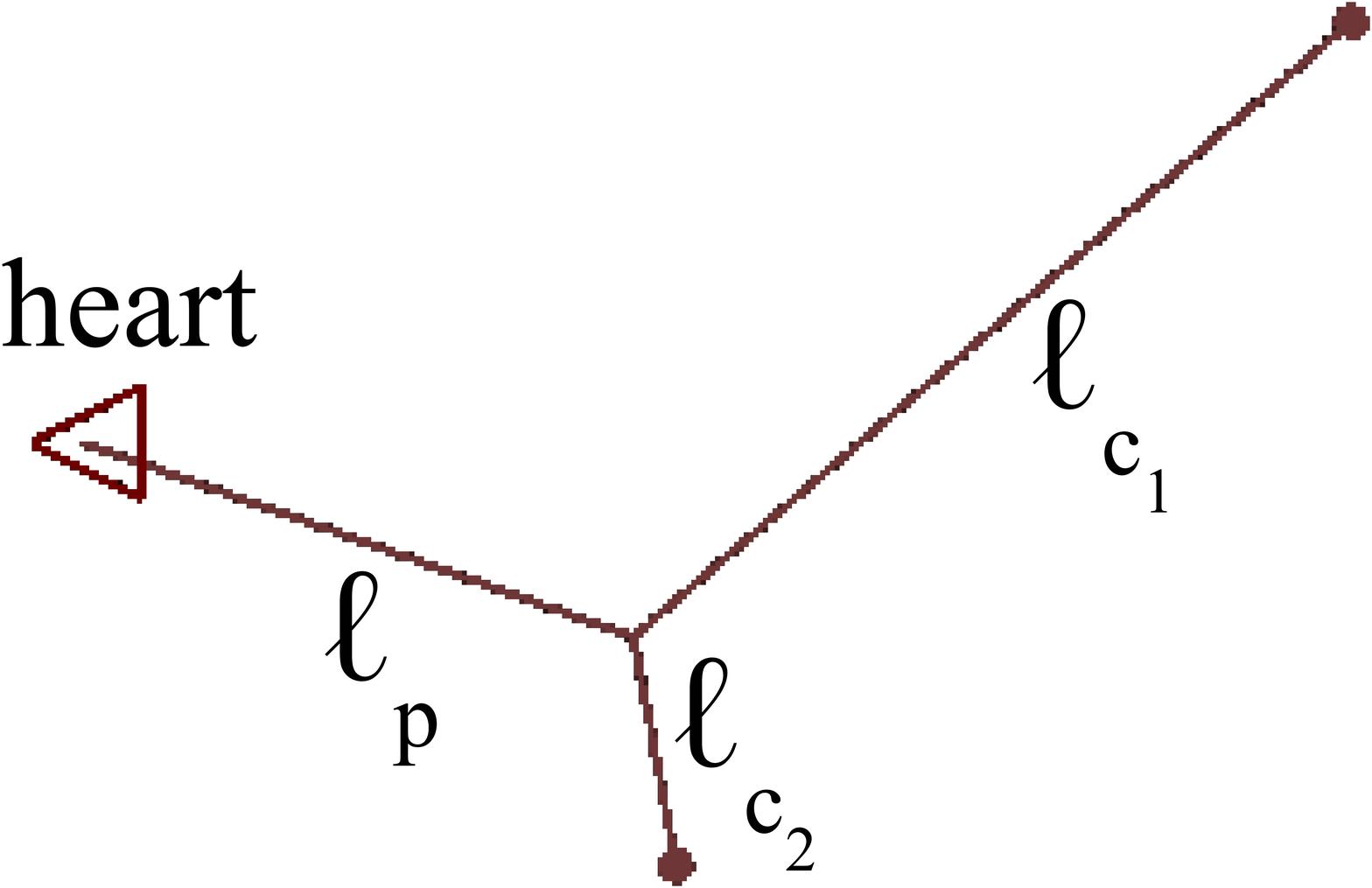}
		\label{labels}
	}
}
&
\\
&
\\[1.2cm]
\cline{1-1}
\parbox{5cm}{
\vspace*{0.5cm}
\centering{
	Local Optimization of Branch Point Position}
}
&
\\
\subfloat[]{
	\includegraphics[width=6cm]{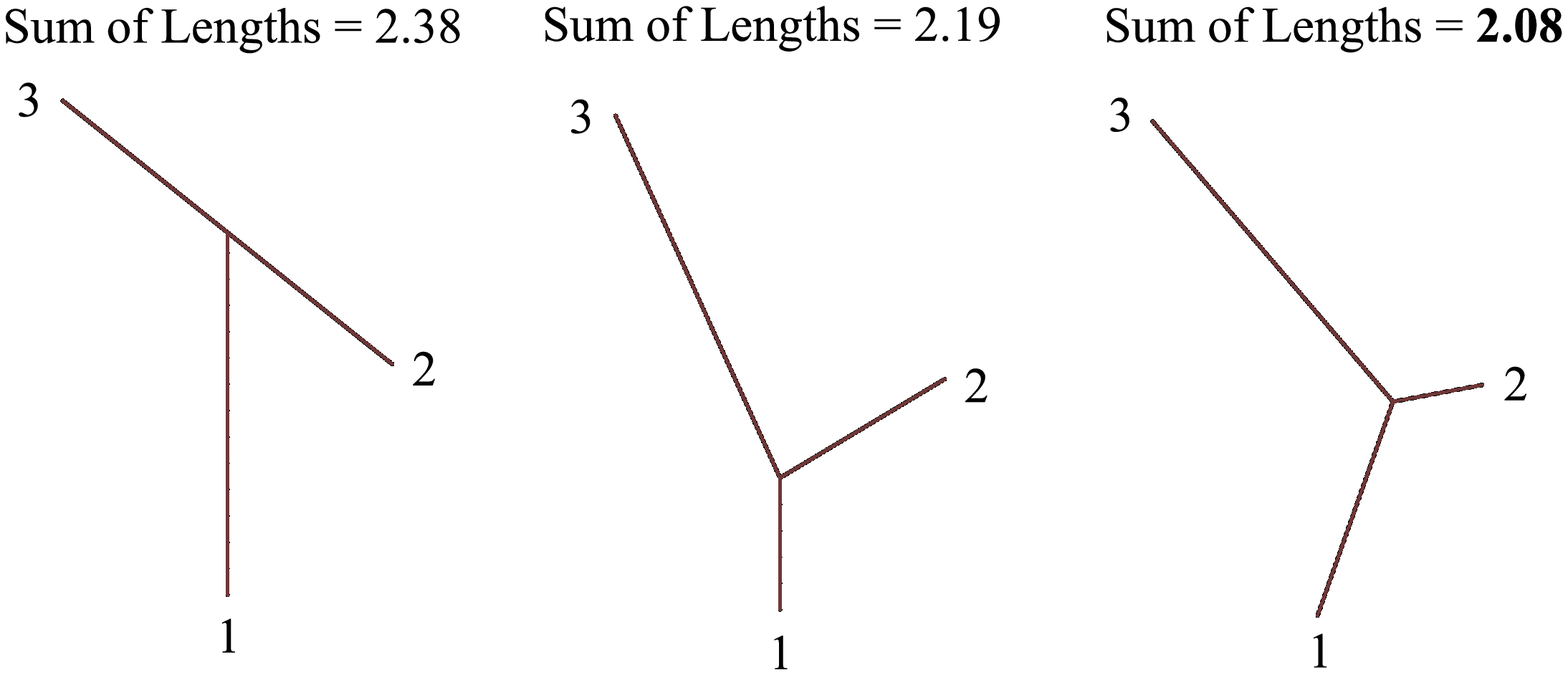}
	\label{fermat}
}
&
\\
\subfloat[]{
	\includegraphics[width=6cm]{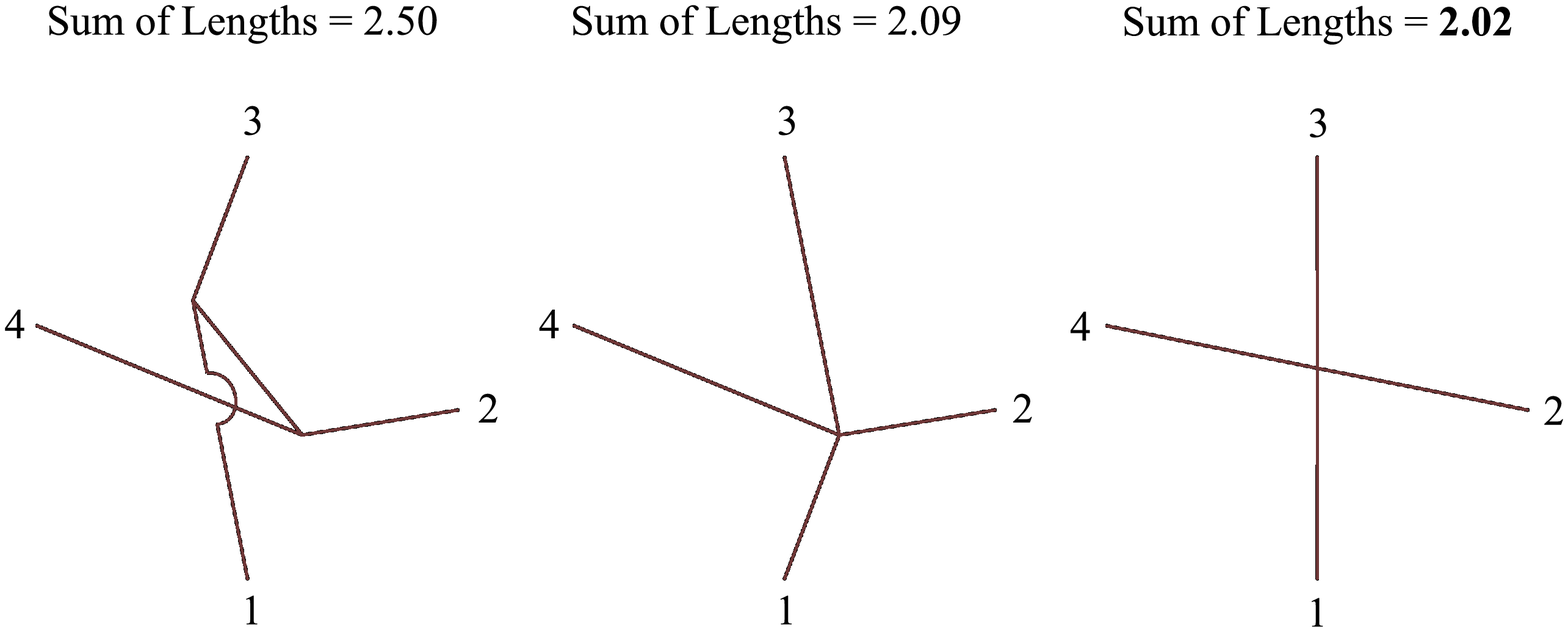}
	\label{geometric_median}
}
&
\multirow{-4}{*}[7.5cm]{
	\vspace*{0.5cm}
	\subfloat[]{
		\includegraphics[width=9cm]{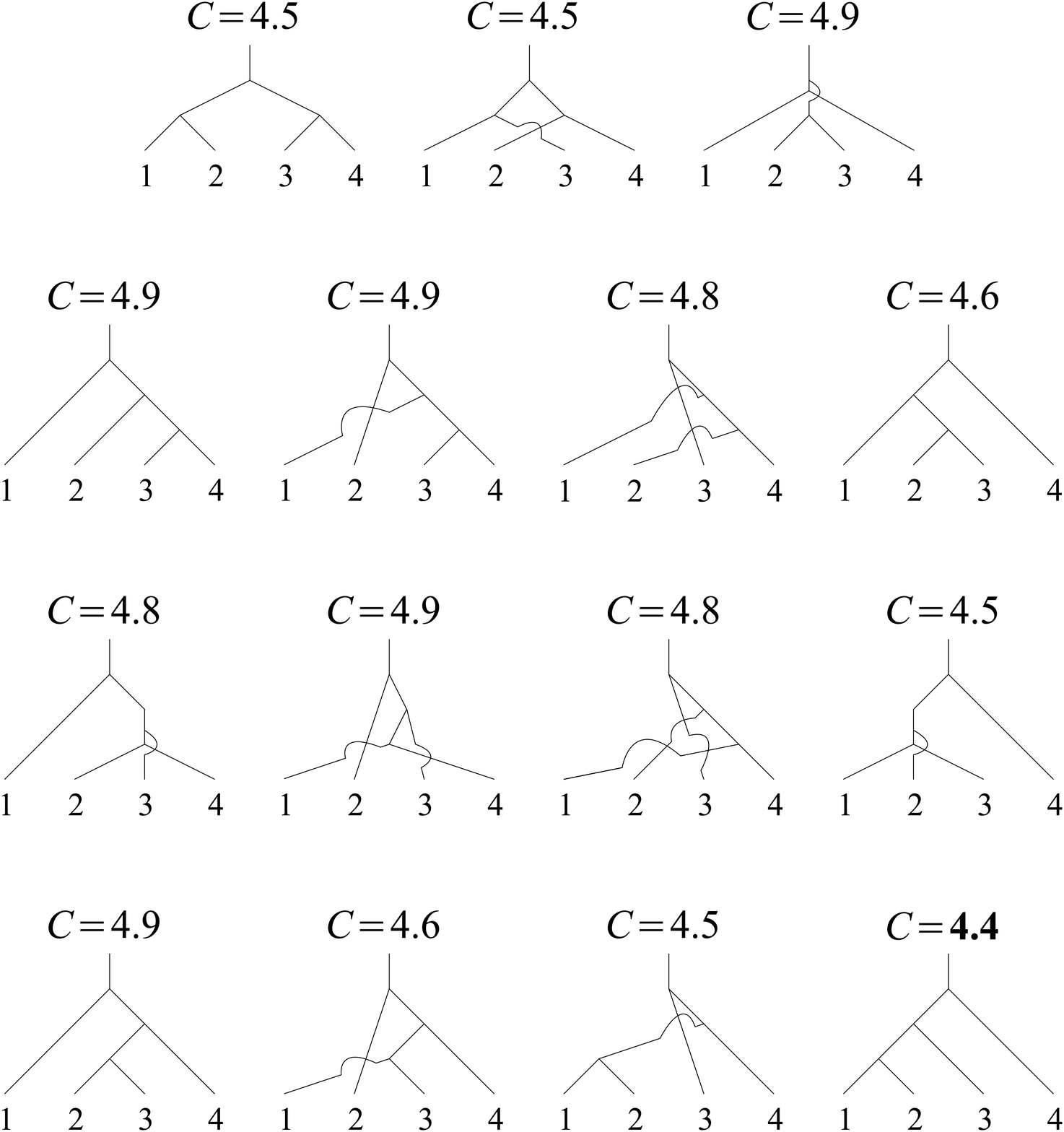}
		\label{tree_search}
	}
}
\end{tabular}
\caption{
	(a)~Schematic of the simplest bifurcating tree network, showing the heart (hollow red triangle) and two service volumes (filled brown circles) with labels for the lengths of each associated segment.
	(b)~Three possible locations for a bifurcation junction.
The rightmost configuration shows the Fermat point of $\bigtriangleup 123$ that minimizes the sum of segment lengths.
	(c)~The two distinct bifurcations (left) collapse to a single trifurcation (center) and set to the geometric median of the four endpoints (right).
	(d)~Comparing $C$ (a measure of some length property of the network) for each of the 15 configurations for four service volumes shows that the bottom right configuration is optimal with respect to $C$.
	The relative order and position of both the tips and branch points in Fig. \ref{tree_search} do not correspond to relative positions in space.
}
\label{basics}
\end{figure*}

We now detail our algorithm for the optimization of the positions of branch points that connect a distribution of service volumes to the heart through a hierarchical network.
Within our algorithm, the position of each branch point depends solely on the location of the adjacent branch points in the network.
Distant vessels affect each other indirectly, but not through any direct long-range process.
Using the limited, local information given by the neighborhood of a branch point, each junction is assigned a uniquely-defined position that minimizes the sum of the Euclidean distances to each neighboring junction.
This is equivalent to the Fermat point of the triangle formed by the two downstream ends of the child vessels and the one upstream end of the parent vessel (see Fig. \ref{fermat}).
The Fermat point of a triangle is a special case of the more general \textit{geometric median}, the unique point that minimizes the sum of distances to an arbitrary number of other fixed points.
We follow the algorithm presented in \cite{Torres12} to avoid errors in determining the geometric median.
Assigning branch point positions as geometric medians effectively minimizes the construction costs for the local network structure.

We construct our networks from simple bifurcations, but using the Fermat point to assign branch point positions can lead to coincident (degenerate) bifurcations, as shown in Fig. \ref{geometric_median}.
Degenerate branch points are consolidated at the geometric median of the upstream endpoint of the parent and three or more downstream endpoints of the associated children segments.
In this way, two degenerate bifurcations become a trifurcation and, more generally, $n$ degenerate bifurcations become a single {$(n + 1)$-furcation}.
Networks that are hierarchically distinct in their bifurcating structure can become identical networks by collapsing bifurcations.
Through exhaustive explorations (described in Sec. \ref{small_networks_subsection}), we find that this marginally reduces the number of possible configurations (see results in Fig. \ref{unique_trees_small}).
Because we have no \textit{a priori} filter to identify which bifurcating trees are redundant, we must consider the entirety of the space of labeled, rooted, bifurcating trees throughout the algorithm to identify sufficiently optimal configurations.
With positions defined in a consistent way, we can now compare the properties of distinct hierarchies to determine which is the best for a particular space-filling strategy.

\section{Selection criteria for biological networks}
\label{quantification_section}

All characteristics of an organism that affect fitness and are heritable are under selection.
A key question is which features of the vascular network are under selection.
Here we define specific fitness measures that are tied to the structure of the network configuration that allows us to rank candidate networks and determine the optimal configuration.

\subsection{Global length properties of space-filling configurations}

Here we introduce a standardized fitness measure that allows us to compare candidate networks for their suitability to transport blood and resources.
Each independent measure for a network's fitness relates to a physical quantity that likely guides the evolution of the cardiovascular system toward a more efficient network.
Specifically, the system's cost in construction material and the maintained volume of the blood relates to the total network length --- the sum of the lengths of all segments.
Competing with this minimization of materials, the dissipation of the heart's pumping power relates to the path lengths between each capillary and the heart.
The power dissipated by smooth (Poiseuille) flow through a segment is directly proportional to the length of the segment \cite{Zamir05}.
In the absence of radius information, reducing the cost of pumping blood is equivalent to reducing the total path lengths that blood travels.

We define these two fitness measures --- one dealing with total network length and the other with individual path lengths --- as
\begin{equation}
L = \sum_{\substack{\rm{all~segments~}i \\ \rm{in~network}}} \ell_i
\label{totalNetworkLength_dfn}
\end{equation}
\begin{equation}
H = \sum_{\substack{\rm{all~paths~}p \\ \rm{in~network}}}~~\sum_{\substack{\rm{all~segments~}i \\ \rm{in~path~}p}} \ell_i
\label{avePathLength_dfn}
\end{equation}
where $N_{tips}$ is the number of distinct service volumes, corresponding to the number of tips and distinct paths.
The generalized total cost function $C$ linearly combines these two measures by their respective weights $C_L$ and $C_ H$:
\begin{equation}
C(C_L, C_H) \equiv C_L L + C_H H.
\label{configuration_fitness_measure_function}
\end{equation}
This cost function connects minimization of material and power dissipation to study optimal networks that are space-filling.

Because an increase in cost corresponds to a decrease in fitness, we place this approach in an evolutionary framework by inverting and scaling our cost function $C$ to be a relative fitness function $F$:
\begin{equation}
F(C_L, C_H) \equiv \frac{C_{min}}{C(C_L, C_H)}
\end{equation}
where $C_{min}$ is the most optimal network under consideration.
By this definition, the optimal configuration has a fitness ${F = 1}$ and less optimal configurations have a fitness ${F < 1}$.

\subsection{Equal Distribution of Resources through Hierarchical Balance}

Because the network tends to exhibit nearly symmetric branching in radius and must distribute resources equally to each capillary in the body, the network hierarchy cannot be overly unbalanced, with one segment having many more tips to supply than its sibling.
In accordance with this argument, empirical data do not show major arteries branching directly into capillaries.
We address this constraint by selecting for networks with more balanced hierarchies.

A hierarchy is better balanced if there are roughly equal numbers of tips supplied downstream by each sibling segment.
Conversely, a hierarchy becomes more poorly balanced as the disparity grows between the number of tips.
In this sense, we define the degree that a hierarchy is unbalanced $U$ as
\begin{equation}
U = 1 - \min_{\substack{\rm{all~sibling}\\\rm{pairs~}(i, j)}}\left\{ \frac{N_{tips}^{(i)}}{N_{tips}^{(j)}} \right\}
\end{equation}
where $N_{tips}^{(i)}$ is the number of distinct downstream service volumes supplied through segment $i$ and segment $j$ is always the sibling with the most downstream tips.
In our algorithm, we select against hierarchies for which the degree of unbalance $U$ is greater than some threshold $U_0$.
We eliminate configurations above this threshold before optimizing branch points and calculating fitness.

\section{Global optimization in the space of hierarchies}
\label{configuration_section}

To determine the optimal hierarchy and its connectivity, we search the space of rooted, labeled, bifurcating trees.
The positions of the branch points are fixed by the process in Sec. \ref{local_optimization_of_positions_subsection}.
The globally optimal network of all configurations maximizes the fitness $F$ while satisfying the space-filling constraint on service volumes.
As an example, the optimal configuration in Fig. \ref{tree_search} corresponds to the hierarchy in the bottom right, where the fitness ${F(1,~0)~=~1}$ includes only total network length $L$ [Eq. (\ref{totalNetworkLength_dfn})], resulting in a Steiner tree \cite{Sankoff75}).

Our exploration of configuration space has many similarities to phylogenetic trees, for which software is available to search through the space of hierarchies \cite{Ronquist12, Guindon2010}. 
Since the available software is not tailored to our specific goals of optimizing space-filling networks, we implement our own algorithms.
Because of the large number of distinct bifurcating rooted trees (that grows factorially with size), efficient search strategies generally focus on regions with greater fitness.
We develop strategies to search through possible configurations and find space-filling networks that best satisfy the general biological constraints from Sec. \ref{quantification_section}.

\subsection{Navigating in the space of hierarchies}
\label{implemented_algorithms_section}

Our numerical technique guides the search by selecting changes that increase configuration fitness.
Making small changes in the branching structure, such as a single swap of two segments in the hierarchy or a regraft of one segment to a spatially near part of the tree, yields new configurations.
Because the change to the hierarchy is small, using the positions of branch points from the previous configuration saves time in optimizing the global positions in the new configuration.

For local swaps in the hierarchy, we exchange a segment with one child (including the associated downstream subtrees) of the segment's sibling (i.e., its nibling).
There are ${2(n~-~2)}$ possible nibling swaps for ${n~(\ge~2)}$ discrete service volumes.
However, nibling swaps do not address changes for segments that are distant in the hierarchy but have small spatial separation.
To account for these changes, we regraft single segments to spatially near branches of the hierarchy.
We limit the search of spatially proximal branch points to those within twice the minimum service volume separation of each other.
This restriction maintains a linear increase in the number of explored regrafts with the number of service volumes, in contrast to the factorial increase that would result from including all possible regrafts.

\subsection{Seed for trajectories: Balanced Hierarchy Construction}

We accelerate the identification of near-optimal networks by choosing an initial configuration that avoids many suboptimal structures (e.g. configurations with many repeated crossings or non-contiguous subtrees).
To improve overall computation time, some approaches explore permutations of pairing tips with service volumes under a constant hierarchy \cite{Weber06}.
Fortunately, the dimensionality of the space for each branch point position never exceeds three in our study, which allows us to construct a favorable configuration directly through spatial partitioning.
Such a favorable configuration avoids less-fit configurations and satisfies the intuitive guidelines that branch points connect nearby subtrees (efficiency by proximity) and that sibling subtrees have similar numbers of service volumes in accordance with symmetric branching in radius.

To ensure the maximal hierarchical balance for the seed, we begin with a single set that contains all terminal service volumes.
This set is then partitioned into two subsets of equal size (or within 1 service volume if the number is odd, which guarantees $U \le 0.5$), using a straight line to define the boundary between the two sets.
When appropriate, this line passes through the geometric center (i.e., the unweighted average position) of the previous set of points and the geometric center of the new subset.
Resulting sets further split into smaller subsets to yield a complete, bifurcating hierarchy.
We refer to this process and the resulting seed as the Balanced Hierarchy Construction (BHC).

\subsection{Efficient search trajectories}

We further accelerate our search by limiting the number of nearby configurations considered at each step.
We accomplish this through a carefully guided greedy search through the space of hierarchies (effectively simulated annealing \cite{Dress87, Salter01} at zero temperature), which often finds a near-optimal configuration.
A greedy strategy offers expedited elimination of configurations that are far from optimal; the algorithm abandons configurations that saturate at a fitness lower than the current most optimal configuration.
Our implementation allows five iterations of the process in Sec. \ref{local_optimization_of_positions_subsection}, then excludes configurations that fail to reduce the cost $C$ [Eq.(\ref{configuration_fitness_measure_function})] by at least 5\% of the remaining difference from the current optimal configuration.
The algorithm with this exclusion scheme successfully identifies near-optimal configurations.

Because the sampling process is not exhaustive, the search through the space of possible hierarchies is not guaranteed to yield a globally optimal configuration.
However, performing a reasonably thorough search as we outline here and conducting several runs from the BHC seed (in our simulations, at least 10 runs) increase the likelihood of identifying a configuration that is near-optimal and shares many of the branching length asymmetries that an optimal configuration exhibits.
With dependable algorithms for determining branch point positions and for exploring the space of possible hierarchies, we can now investigate the length properties of space-filling networks under several basic space-filling strategies.

\section{Results and Analysis}
\label{results_section}

We now present the results of optimized networks and of the analysis on real vascular networks, including the properties of the most optimal networks.
To build intuition about the space of hierarchies, we first explore the space exhaustively for small networks and establish the distinct patterns that the two optimizations $L$ and $H$ produce.
In comparing optimal configurations with observations of real systems, we find better agreement by enforcing a constraint on the degree of unbalance $U$ in the hierarchies of candidate configurations.

\subsection{Exhaustive search for small networks}
\label{small_networks_subsection}

To become more familiar with the landscape of possible configurations, we exhaustively explore the space of hierarchies and pairings for networks that are small enough to quickly yield comprehensive results for a single realization of fixed service volumes.
We collapse and reorganize the higher-dimensional space of branch point swaps into a single dimension by ranking each configuration based on the fitness $F$.
This reorganization involves a normalization of rank so that the fittest configuration occurs at 0 and the least fit occurs at 1.
\begin{figure*}[ht]
\vspace*{0.5cm}
\centering
\begin{tabular}[t]{c@{\hskip 2cm}c}
\parbox{6cm}{
	\centering{
		Fitness Landscape $F(1, 0)$
	}
}
&
\parbox{6cm}{
	\centering{
		Fitness Landscape $F(1, 9)$
	}
}
\\
\subfloat[]{
	\includegraphics[width=7.3cm]{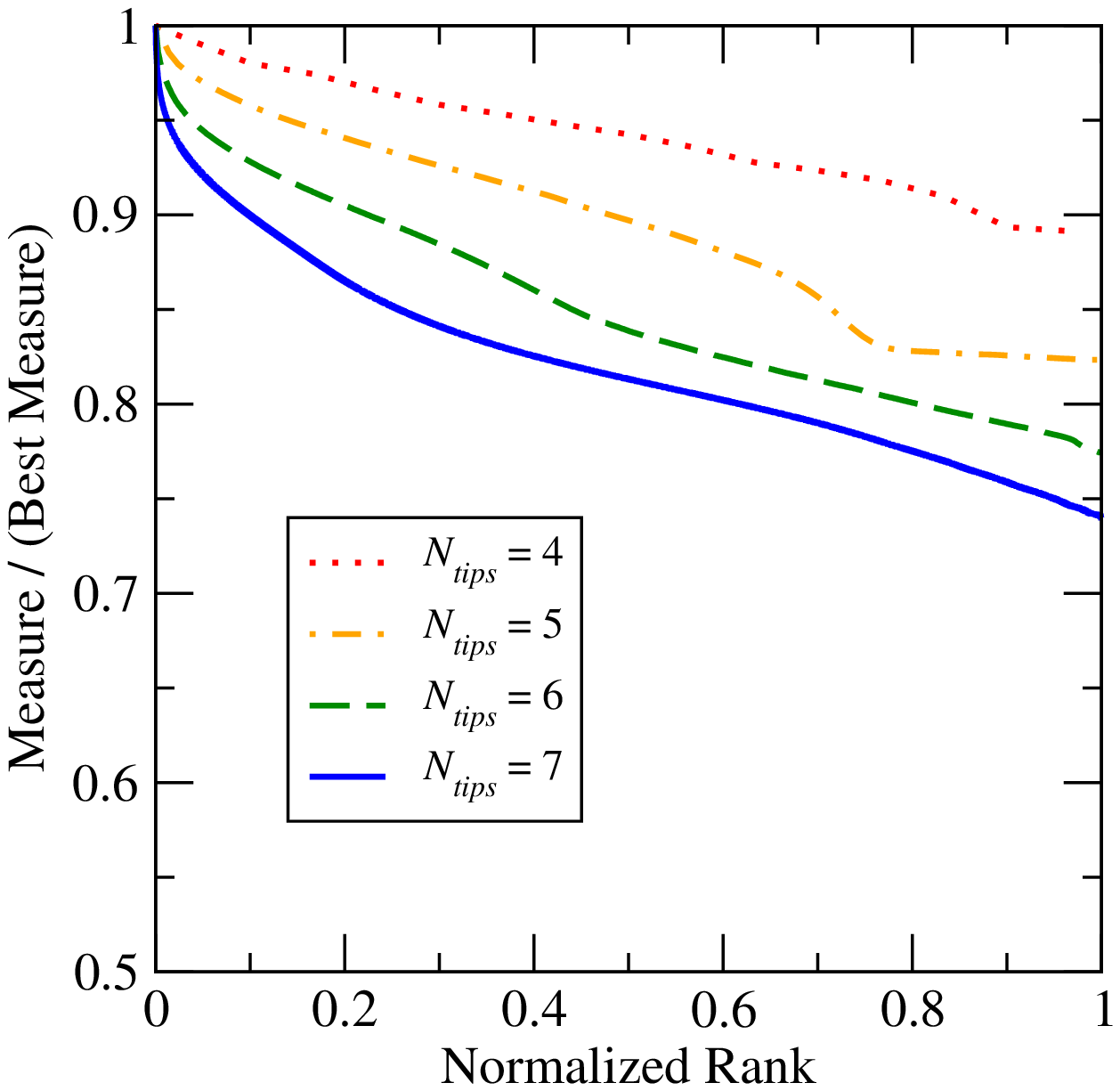}
	\label{exhaustive_rank_by_measure}
}
&
\subfloat[]{
	\includegraphics[width=7.3cm]{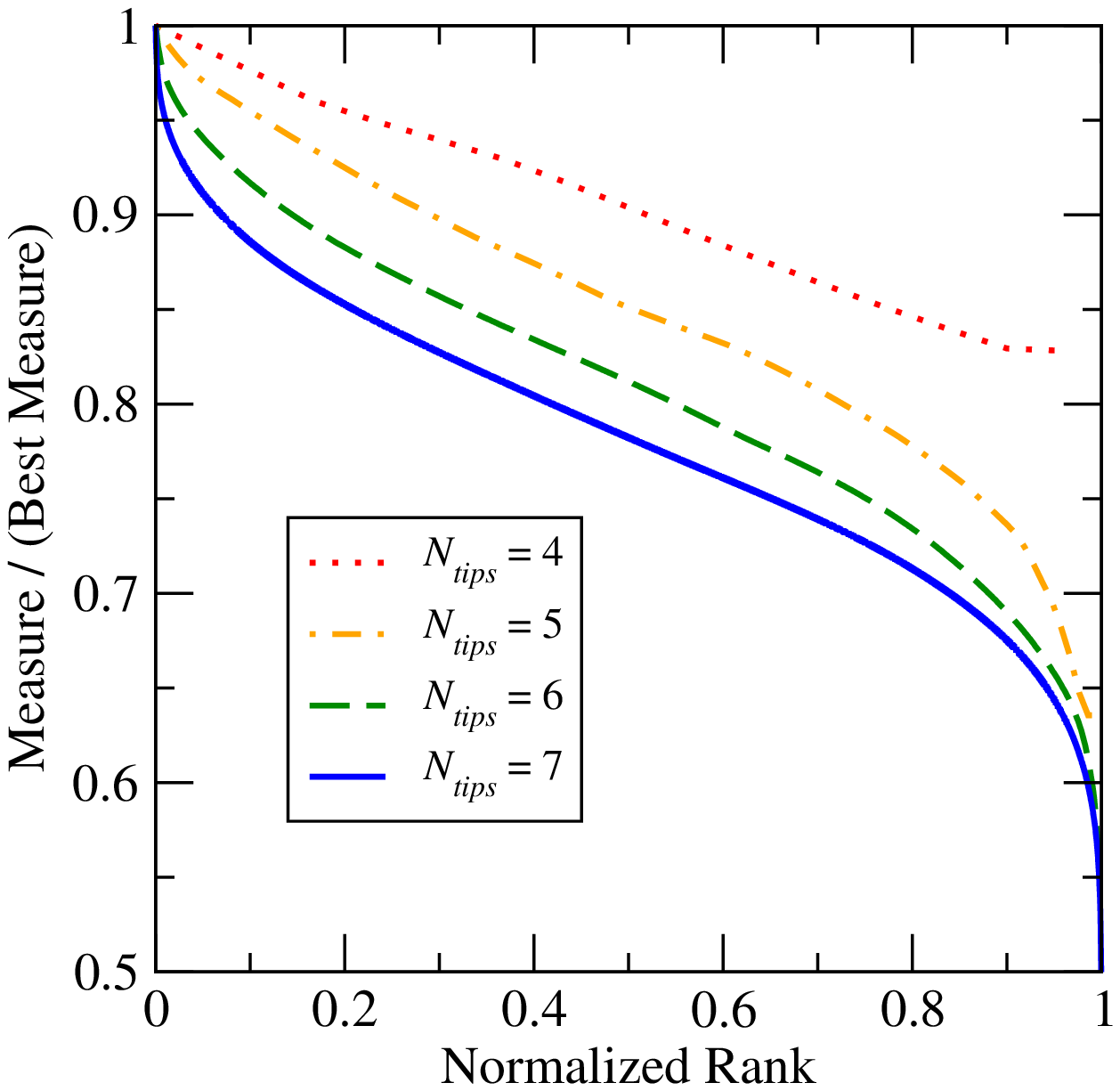}
	\label{exhaustive_rank_by_measure_c19}
}
\end{tabular}
\caption{
	(Color)
	Ranked landscapes for unique consolidated configurations from exhaustive exploration of bifurcating trees of fixed area and fixed $N_{tips}$
	(a)~Exhaustive landscapes for fitness based only on total network length $L$ ${[F(1,~0)]}$.
	(b)~Exhaustive landscapes for fitness based total network length $L$ and average path length $H$ ${[F(1,~9)]}$.
	We choose the weights ${(C_L,~C_H)~=~(1, 9)}$ so that the contribution from $H$ is not dominated by the contribution from $L$.
}
\label{small_network_landscapes}
\end{figure*}
Rescaling the rank is necessary even for networks of the same size and shape because different realizations may have different numbers of unique configurations after consolidating degenerate bifurcations (mentioned in Sec. \ref{local_optimization_of_positions_subsection}), despite having the same number of service volumes.
Larger networks tend to have greater range with respect to both costs in $L$ and $H$.

The minimum distance between each service volume is constant for each of the networks that constitute the ensemble of realizations for the curves in Fig. \ref{exhaustive_rank_by_measure}.
Each curve represents the average fitness (relative to the fittest configuration for the particular set of capillary positions) over an ensemble of networks with a fixed number of tips and constant total area.
In generating the ensemble, we exclude those that arise with a different number of tips than what is desired until we accumulate 1000 configurations of the target size.
Across curves, the total area increases to produce networks with more service volumes more frequently.

One might expect a large set of similarly fit, near-optimal networks, which would be represented by a plateau near the optimum.
However, the sharp descent away from the optimal configuration in Fig. \ref{exhaustive_rank_by_measure} indicates that there are few configurations that are near-optimal.
From an evolutionary perspective, this implies that the vascular networks of organisms are under strong selection. 
Furthermore, the slope near the optimum becomes steeper as more service volumes are introduced, so that the best configurations become more distinct from other possibilities as the number of service volumes grows.
Considering the very large number of service volumes in real organisms, this again indicates that real vascular networks are under strong selection pressures for space-filling and efficiency.

\begin{figure}[ht]
\centering
\vspace*{0.5cm}
\begin{tabular}[t]{c@{\hskip 1cm}c}
\parbox{3.6cm}{
	\centering{
		Global Optimization for Total Network Length $L$ $[F(1, 0)]$
	}
}
&
\parbox{3.6cm}{
	\centering{
		Global Optimization for Average Path Length $H$ $[F(0, 1)]$
	}
}
\\
\subfloat[]{
	\includegraphics[width=3.6cm]{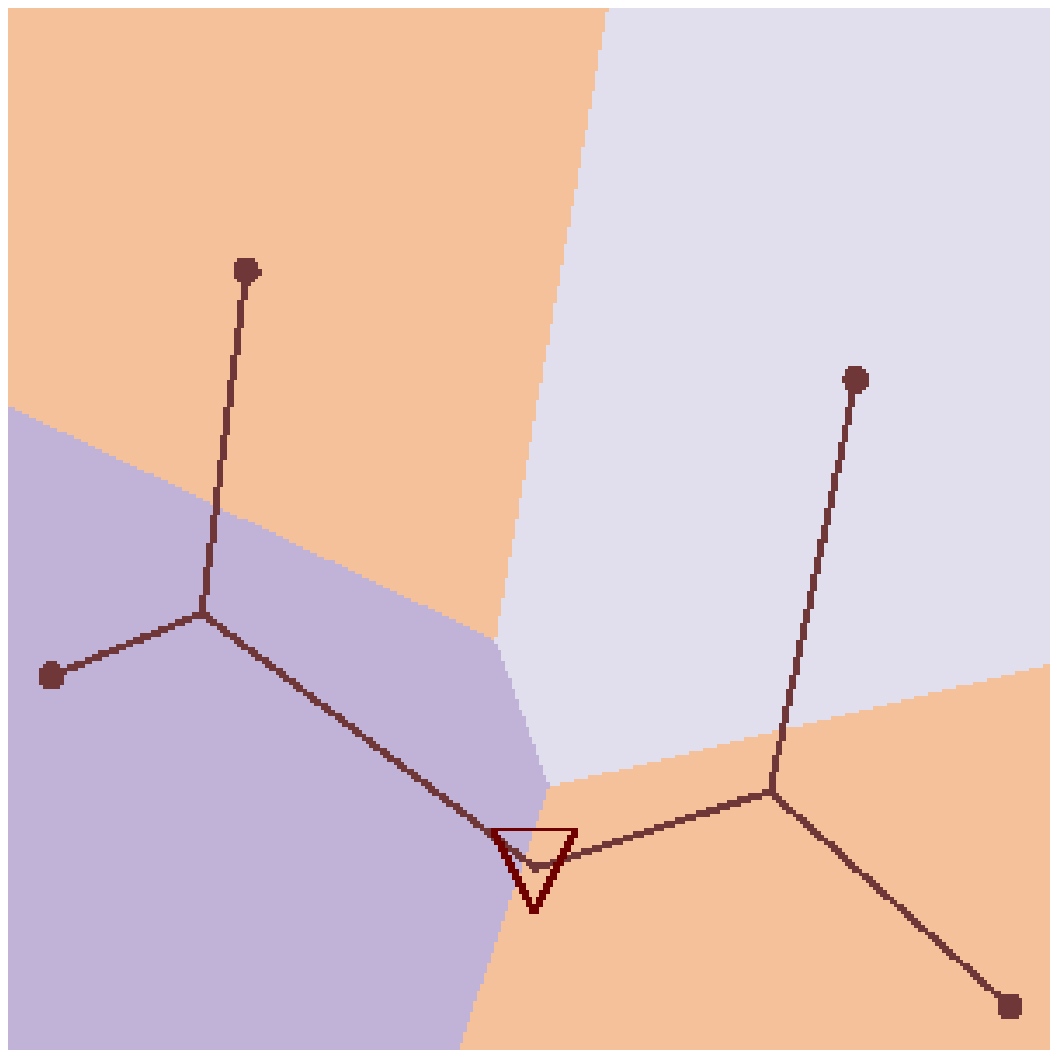}
	\label{total_length_bifur}
}
&
\subfloat[]{
	\includegraphics[width=3.6cm]{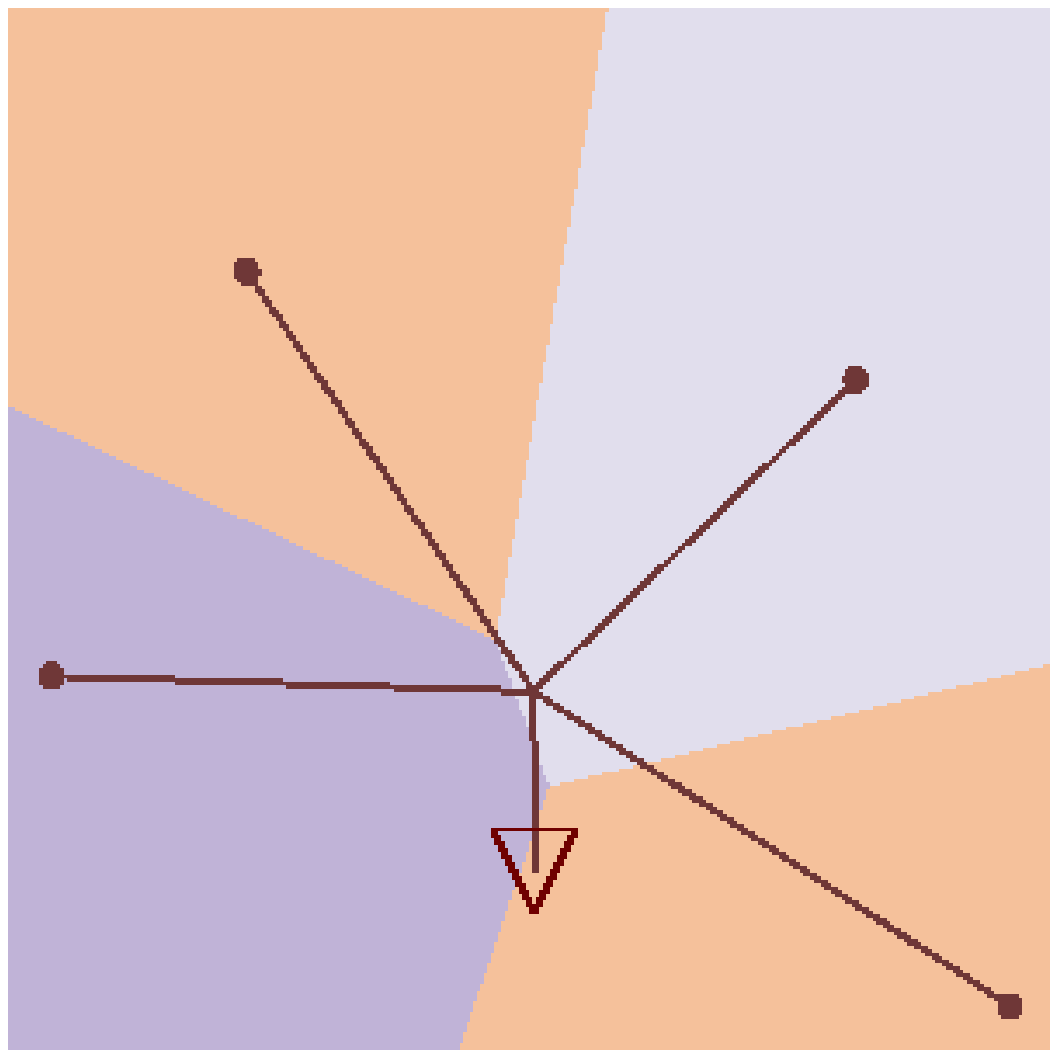}
	\label{path_length_hub}
}
\end{tabular}
\caption{
	(Color)
	Two classes of networks: (a)~The optimal configuration that minimizes total network length {[$L$ in Eq. (\ref{totalNetworkLength_dfn})]} of the 15 possible trees (corresponding to Fig. \ref{tree_search}) consists only of bifurcations.
	(b)~The optimal configuration that minimizes average path length between each service volume and the heart {[$H$ from Eq. (\ref{avePathLength_dfn})]} of the 15 possible trees consists of a single hub.
	The regions of varying background color define the Voronoi cells corresponding to individual service volumes.
}
\label{simpleNetworks}
\end{figure}

Optimal networks that have no constraint on hierarchical balance fall into two general classifications depending on the relative weights of total network length $L$ and average path length $H$ in the fitness measure $F$.
As shown in the simple examples of Fig. \ref{simpleNetworks}, network fitness measures that are weighted to minimize $L$ yield bifurcating trees, while measures that are weighted to minimize $H$ yield hubs.
Bifurcating trees better correspond to real networks, suggesting that  total network length $L$ plays a larger role than average path length $H$.
Since a single hub is not observed (and not expected from material costs) in real systems, we do not consider configurations that ignore total network length $L$.
However, optimizing only for $L$ leads to meandering, bifurcating paths, which become shorter and more direct by including both costs ($L$ and $H$).
Furthermore, additional global information is necessary to directly minimize path lengths than the local environment that we consider in Sec. \ref{local_optimization_of_positions_subsection} --- specifically, the context of the entire path.
This means that our analysis is best suited for optimality that always includes a significant contribution from total network length $L$ and a weaker contribution from average path length $H$.

\subsection{Trajectories for sampling larger networks}
\label{larger_networks_subsection}

With better intuition about the space of hierarchies from small networks, we now explore the space for larger networks with more service volumes.
The branching properties of larger networks give more applicable results to connect particular space-filling strategies with the observations of real cardiovascular systems.
We first summarize the properties of optimized networks without any constraint on hierarchical balance (${U_0~=~1}$).

Because the search through the space of hierarchies is not exhaustive for large networks, we cannot show ranked landscapes averaged over ensembles with different service volume positions as we did for small networks.
Instead, we show landscapes from a single realization of service volume positions that come from an ensemble of trajectories that start with the BHC configuration and end at a local optimum  (Fig. \ref{guess_and_random_runs} in App. \ref{size_and_shape_section}).
\begin{figure}[ht]
\centering
\vspace*{1cm}
\includegraphics[width=8.3cm]{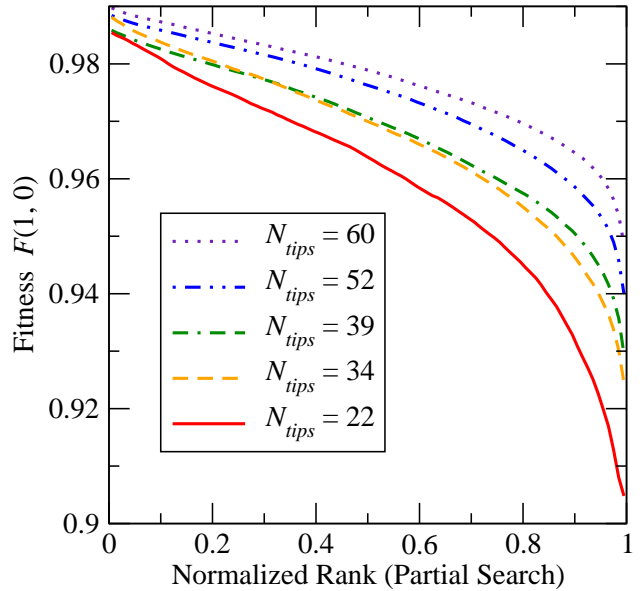}
\label{large_rank_by_measure}
\caption{
	(Color)
	Average fitness landscapes for total network length $L$ ${[F(1,~0)]}$ over 100 trajectories for a single network of each size.
}
\label{large_network_measures}
\end{figure}
The greedy algorithm samples fewer less-fit configurations, yielding a shallower slope near the optimum than the exhaustively explored landscapes in Fig. \ref{exhaustive_rank_by_measure}.
Since the starting point of the search (the BHC configuration) is already favorable, we expect that the worst-ranked configuration of the partial search is already very near optimal.

\begin{figure*}[ht]
\captionsetup[subfigure]{labelformat=empty}
\vspace*{0.5cm}
\centering
\begin{tabular}{ccc}
\parbox{5cm}{
	\centering{BHC Seed}
}
&
\parbox{5cm}{
	\centering{Total Network Length $L$ $[{F(1,~0)}]$}
}
&
\parbox{5cm}{	\centering{Total Network Length $L$ and Mean Path Length $H$ $[{F(1,~9)]}$}
}
\\
\subfloat[]{
	\includegraphics[width=5.3cm]{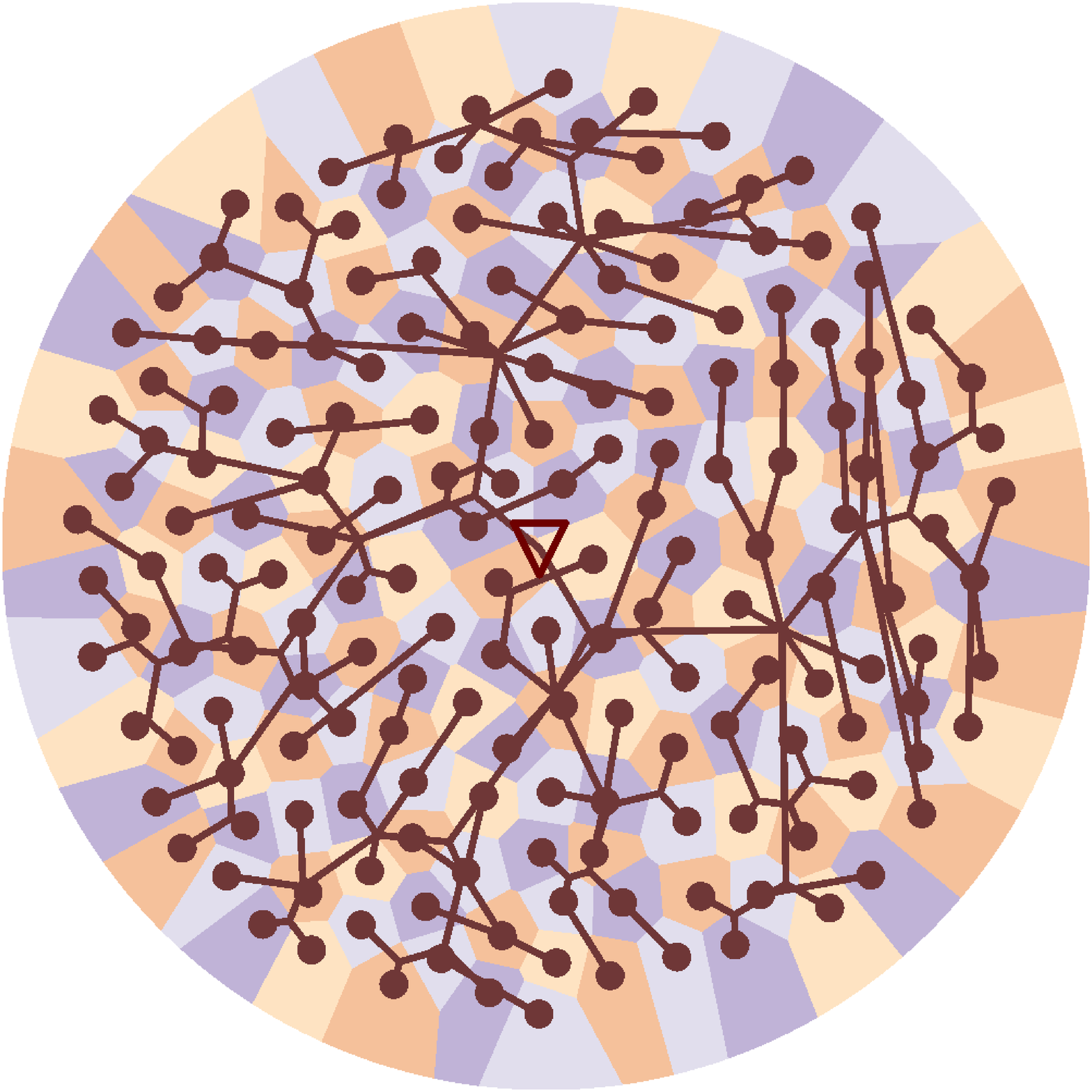}
	\label{circ_bhc}
}
&
\subfloat[]{
	\includegraphics[width=5.3cm]{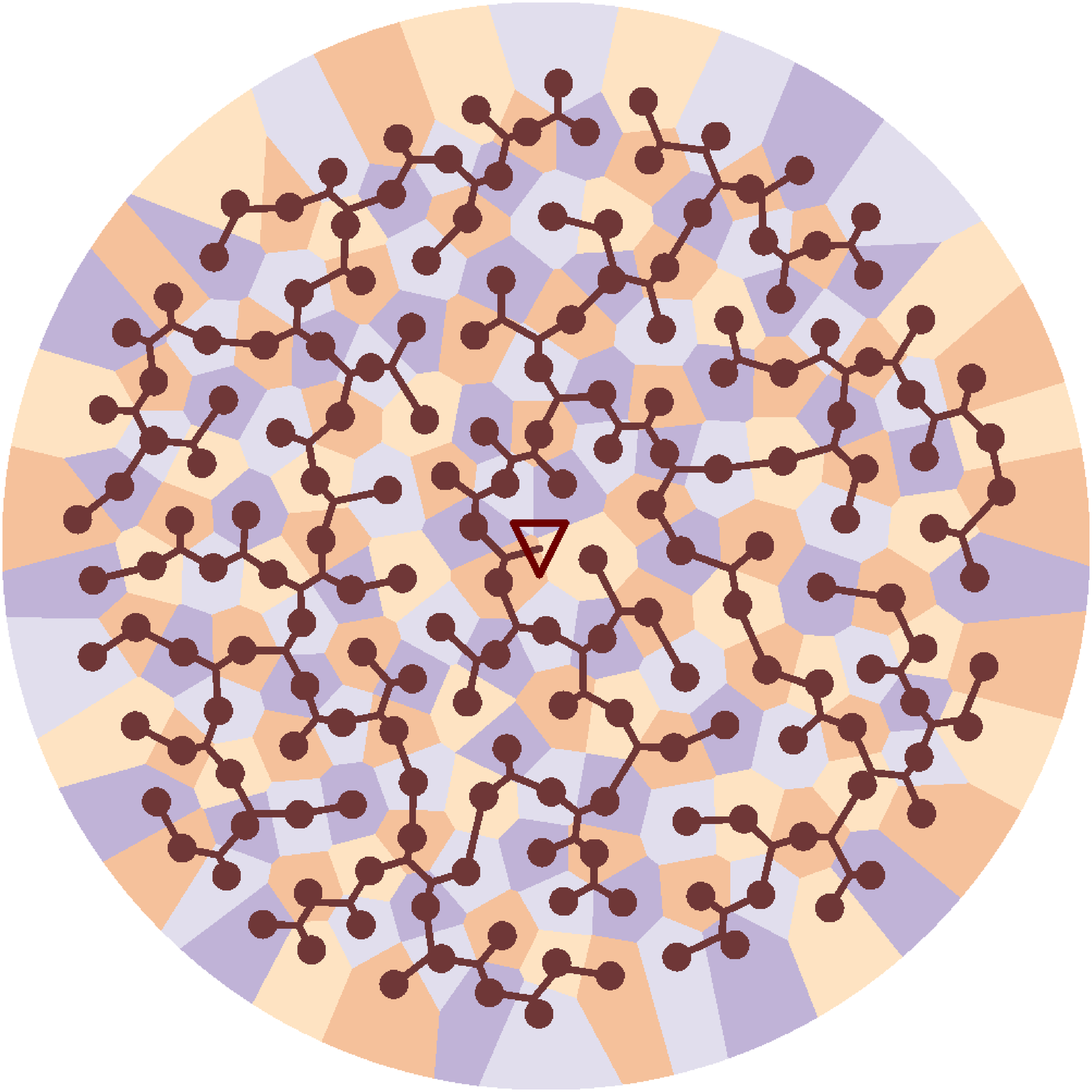}
	\label{circ_c100}
}
&
\subfloat[]{
	\includegraphics[width=5.3cm]{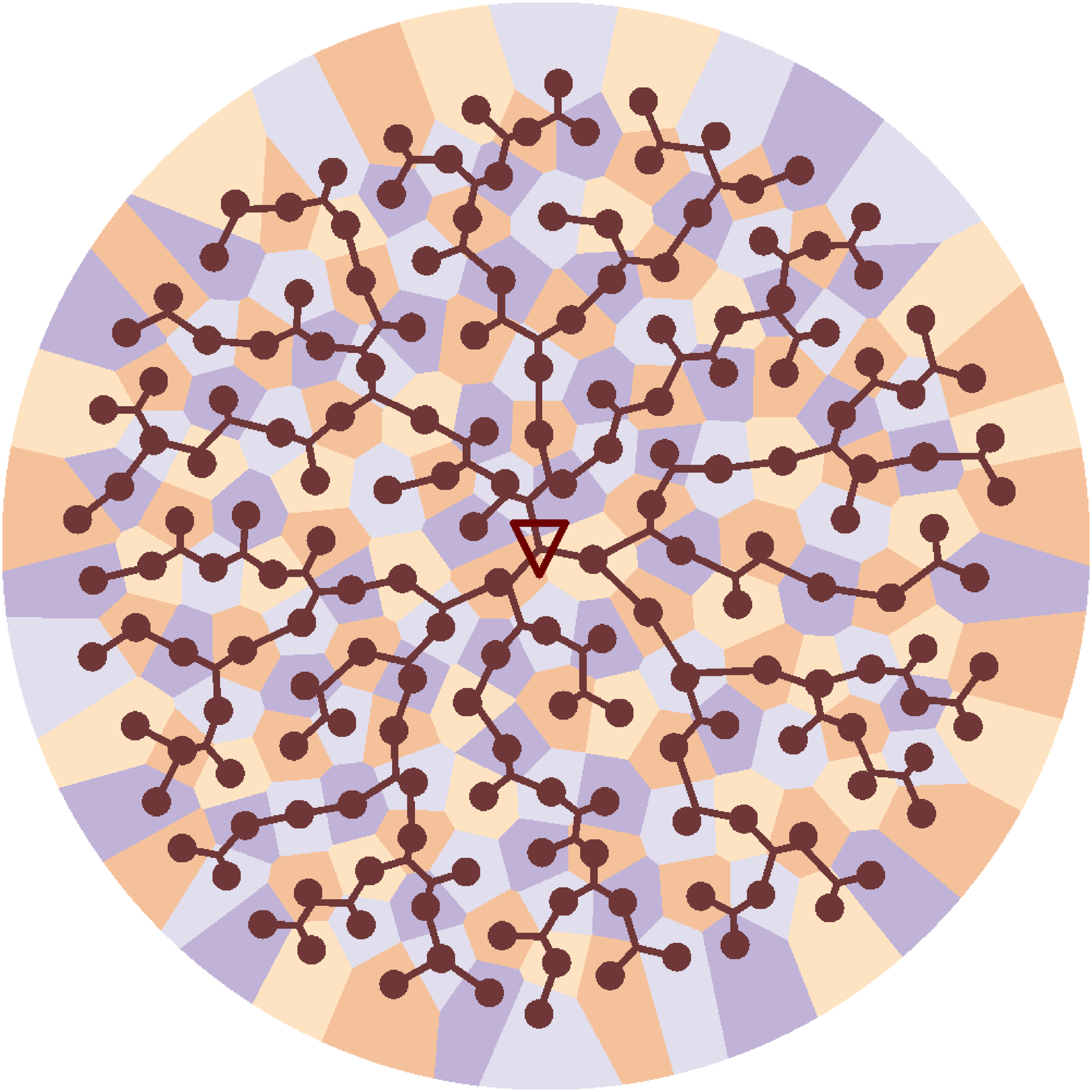}
	\label{circ_c190}
}
\end{tabular}
\caption{
	(Color)
	Optimal configurations for two fitness measures.
	We choose the weights ${(C_L,~C_H)~=~(1, 9)}$ so that the contribution from $H$ is not dominated by the contribution from $L$.
}
\label{large_optimal_networks_circ}
\end{figure*}

Searches through the space of hierarchies and the properties of optimal configurations do not vary with different convex body shapes.
Fig. \ref{large_optimal_networks_circ} shows example optimized networks for a maximally symmetric body shape (see App. \ref{size_and_shape_section} for other shapes).
The general trends of long, meandering paths for solely minimizing total network length $L$ and of more direct paths when including $H$ are consistent across both isotropic, circular areas and elongated, rectangular areas.

\begin{figure}[ht]
\vspace*{1cm}
\centering
\includegraphics[width=7.3cm]{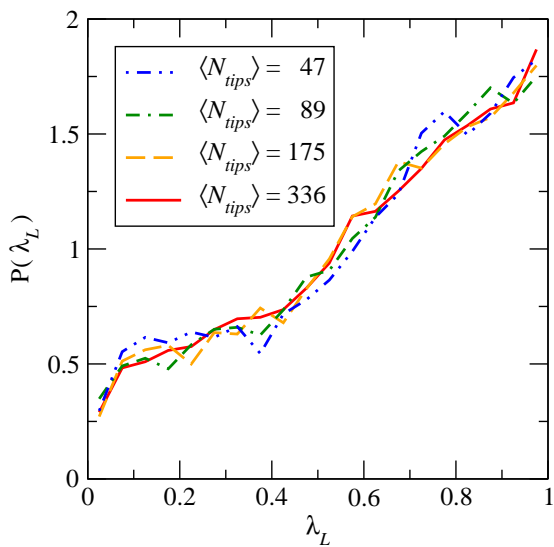}
\label{no_constraint_circ_sizes_optimal_lambda_L}
\caption{
	(Color)
	Distributions of $\lambda_L$ for several sizes of circular areas, averaged over 100 realizations of service volume distributions and optimized solely for total network length $L$ ${[F(1,~0)]}$.
	Local length asymmetry between siblings skews toward symmetry for optimal networks with no hierarchical balance.
}
\label{no_constraint_optimal_lambda_L}
\end{figure}

To characterize branching features of these large configurations, we quantify the asymmetric branching attributes with the two ratios
\begin{equation}
\lambda_L = \frac{\ell_{c_1}}{\ell_{c_2}}
\label{lambdaL_dfn}
\end{equation}
\begin{equation}
\lambda_R = \frac{r_{c_1}}{r_{c_2}}
\label{lambdaR_dfn}
\end{equation}
choosing $\ell_{c_1} \le \ell_{c_2}$ for the lengths of child 1 and 2 and $r_{c_1} \le r_{c_2}$ for the radii (shown in Fig. \ref{labels}).
Note that perfect symmetry corresponds to ${\lambda_L~=~\lambda_R~=~1}$ and smaller values of $\lambda_L$ and $\lambda_R$ correspond to more asymmetric branching.
Distributions for the branching asymmetry ratio in length $\lambda_L$ for various sizes and shapes are shown in Fig. \ref{no_constraint_optimal_lambda_L}.
Branch points for which one of the child segments does not exist because of degeneracy with a service volume center do not contribute to the distribution for $\lambda_L$.
There is little change in the features of the distribution of $\lambda_L$ across different sizes of networks with $U_0 = 1$.
This trend persists for both isotropic and anisotropic enclosing shapes (as Fig. \ref{no_constraint_optimal_lambda_L} in App. \ref{size_and_shape_section} shows).
A summary for the cross-generational length ratio is given in App. \ref{gamma_section}.

Many branch points in these networks coincide with a service volume, predicting large trunks that feed capillaries directly.
Similar results appear in the study of flow through a dynamic, adaptive network \cite{HuCai13}.
However, such a trend does not agree with the empirical data.
Although there is asymmetry in adjacent segments at branch points and a lack of strict balance in the hierarchy along different paths, we observe that large arteries do not branch directly to capillaries and arrive at the same expectation from the dynamics of blood flow.
The major qualitative distinction between the BHC and the optimized configurations with ${U_0~=~1}$ is that the BHC is a network with a balanced hierarchy.
Upon inspection of empirical data in Sec. \ref{empirical_comparison_section}, we find that the branching length asymmetries for the BHC configuration (given in App. \ref{size_and_shape_section}) motivate an additional constraint on hierarchical balance during the search through the space of hierarchies.

\subsection{Comparison of optimized networks with empirical data}
\label{empirical_comparison_section}

The results in Sec. \ref{larger_networks_subsection} show that optimization for total network length or average path length with no constraint on hierarchical balance leads to distributions of asymmetry in sibling vessel length that skew toward symmetry ($\lambda_L \approx 1$).
We now present the analysis of $\lambda_L$ that characterizes the local length asymmetries at branch points for real and optimized networks.
From this analysis, we explore how limiting the degree of unbalance $U$ in an optimal artificial network yields asymmetries that better match biological networks.

\subsubsection{Asymmetric vessel length distributions of real networks}

\begin{figure*}[ht]
\centering
\vspace*{0.5cm}
\begin{tabular}[t]{c@{\hskip 2cm}c}
\parbox{5.5cm}{
	\centering{
		Sibling Radius Ratio ($\lambda_R$)
	}
}
&
\parbox{5.5cm}{
	\centering{
		Sibling Length Ratio ($\lambda_L$)
	}
}
\\
\subfloat[]{
	\includegraphics[clip, width=7.3cm]{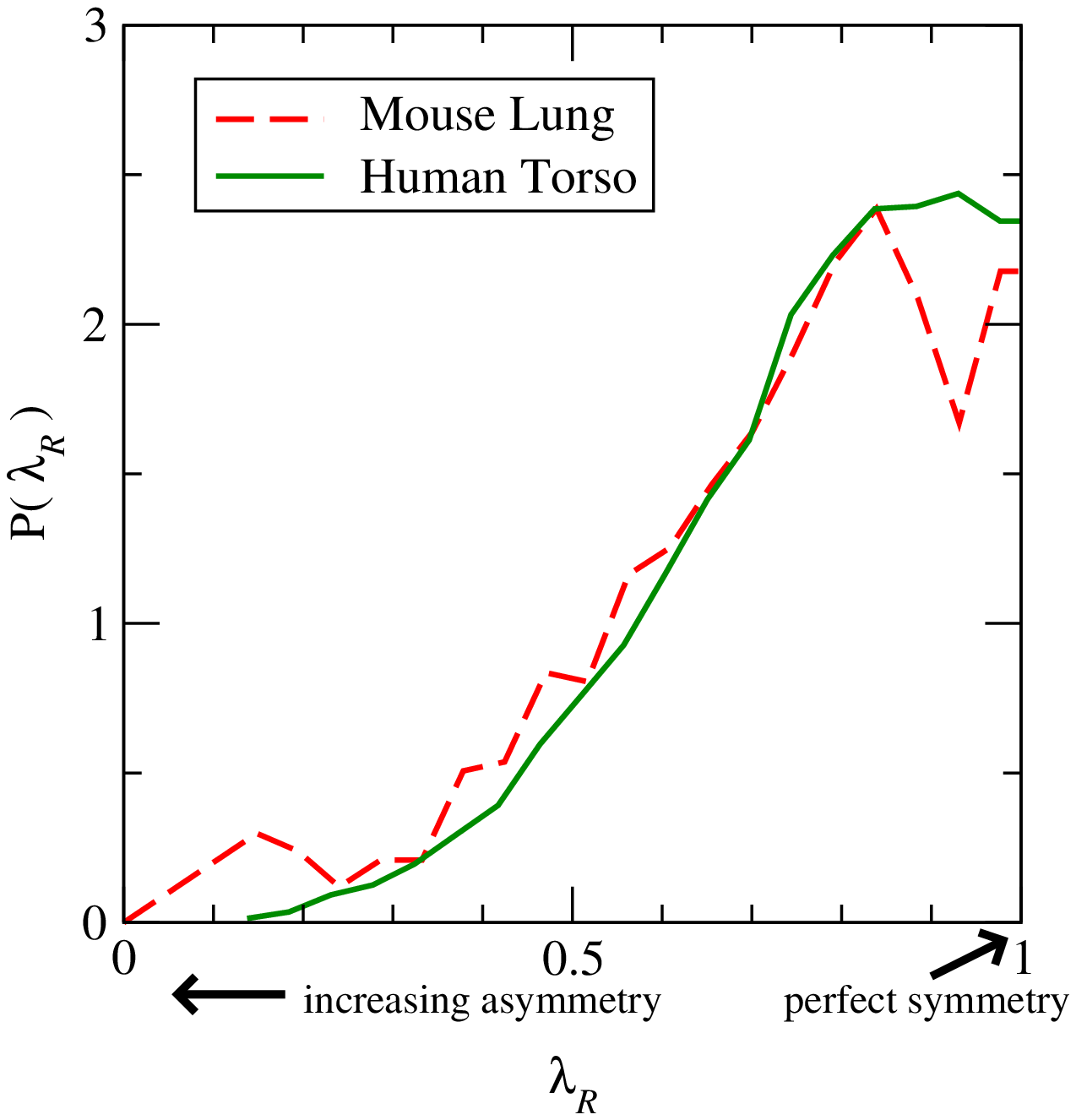}
	\label{lambdaR_dicom}
}
&
\subfloat[]{
	\includegraphics[clip, width=7.3cm]{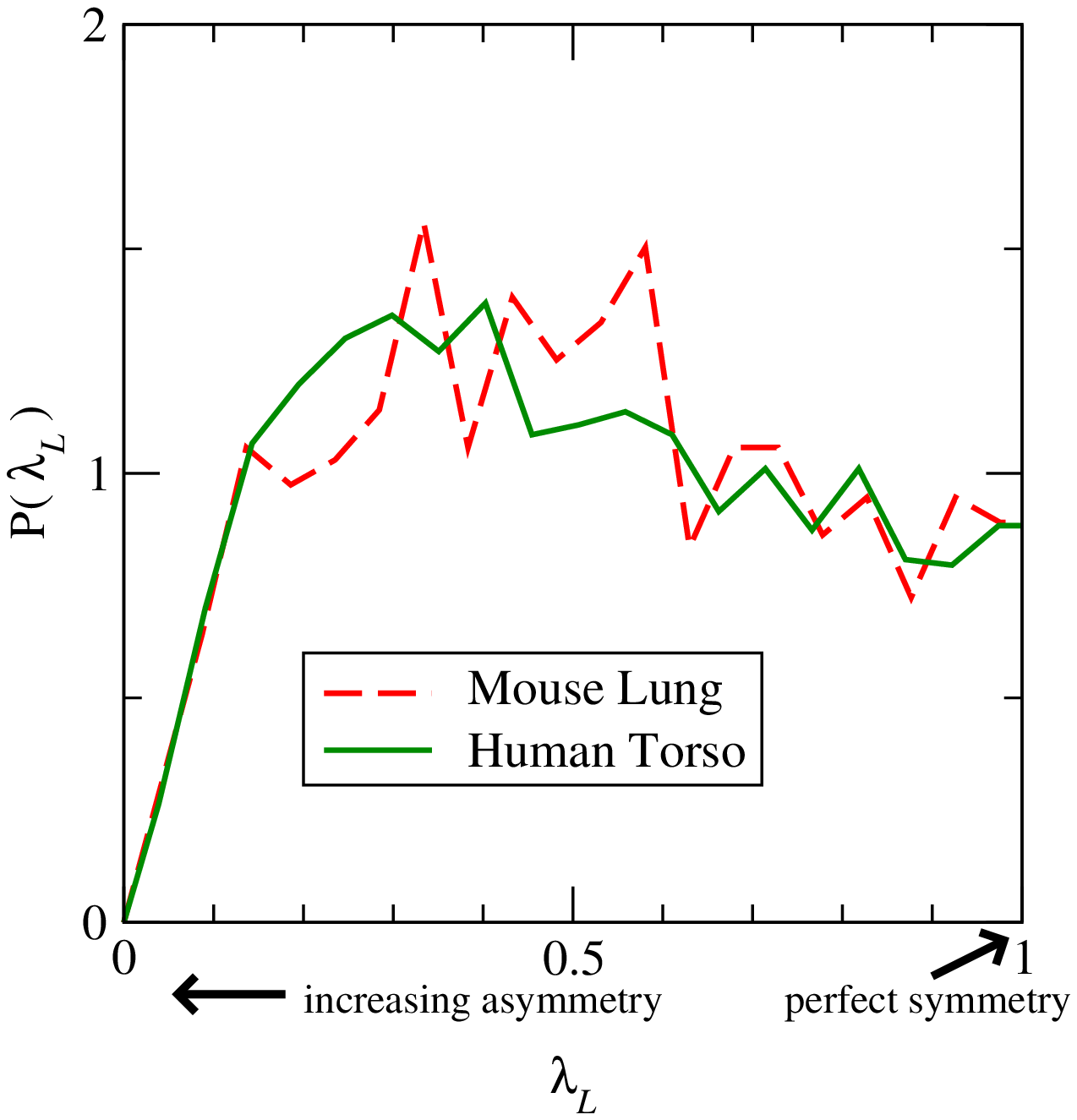}
	\label{lambdaL_dicom}
}
\end{tabular}
\caption{
	(Color)
	Observed radius and length branching asymmetry ratios [Eqs. (\ref{lambdaL_dfn}) and (\ref{lambdaR_dfn}), respectively] in mouse lung and human torsos.
	(a)~Radii ratios are skewed toward symmetry (${\lambda_R \approx 1}$), although they are not always perfectly symmetric.
	(b)~Length ratios are not skewed toward symmetry (many ratios have ${\lambda_L~<~1}$), contrary to symmetric models.
}
\label{observed_asymmetry}
\end{figure*}

We analyze MRI images of the human head and torso as well as micro tomography images from wild-type mouse lung.
Both data sets break from strict symmetry.
As shown in Fig. \ref{observed_asymmetry}, the network-wide distribution for $\lambda_R$ is skewed toward symmetry (${\lambda_R\approx~1}$), while the distribution for $\lambda_L$ is more uniform, representing a greater contribution from very asymmetric branching (${\lambda_L~<~1}$).
These results are representative of general features for length distributions in real biological networks.
The fact that the optimized networks in Sec. \ref{larger_networks_subsection} do not exhibit a similar distribution for $\lambda_L$ signals that important biological factors are missing.
Because of the skew toward symmetry in sibling segment radii, we limit the hierarchical unbalance of optimized networks in Sec. \ref{hierarchical_constraint_section}.

\subsubsection{Degree of balance necessary to match biological networks}
\label{hierarchical_constraint_section}

\begin{figure}[ht]
\captionsetup[subfigure]{labelformat=empty}
\centering
\vspace*{0.5cm}
\begin{tabular}{c|cc}
{}
&
\parbox{3.6cm}{
	\centering{Total Network Length $L$ ${[F(1,~0)]}$}
}
&
\parbox{3.6cm}{
	\centering{Total Network Length $L$ and Mean Path Length $H$ ${[F(1,~9)]}$}
}
\\
\parbox{1.9cm}{\raggedleft $U_0 = 1.0$}
&
\subfloat[]{
	\includegraphics[width=3cm]{singleNetOptRec_c1.0-0.0-0.0_hierSym0.000_side10.1circ_00001_optGraft.eps}
	\label{p0_c100}
}
&
\subfloat[]{
	\includegraphics[width=3cm]{singleNetOptRec_c1.0-9.0-0.0_hierSym0.000_side10.1circ_00004_optGraft.eps}
	\label{p0_c190}
}
\\
\parbox{1.9cm}{\centering{$U_0 = 0.9$}}
&
\subfloat[]{
	\includegraphics[width=3cm]{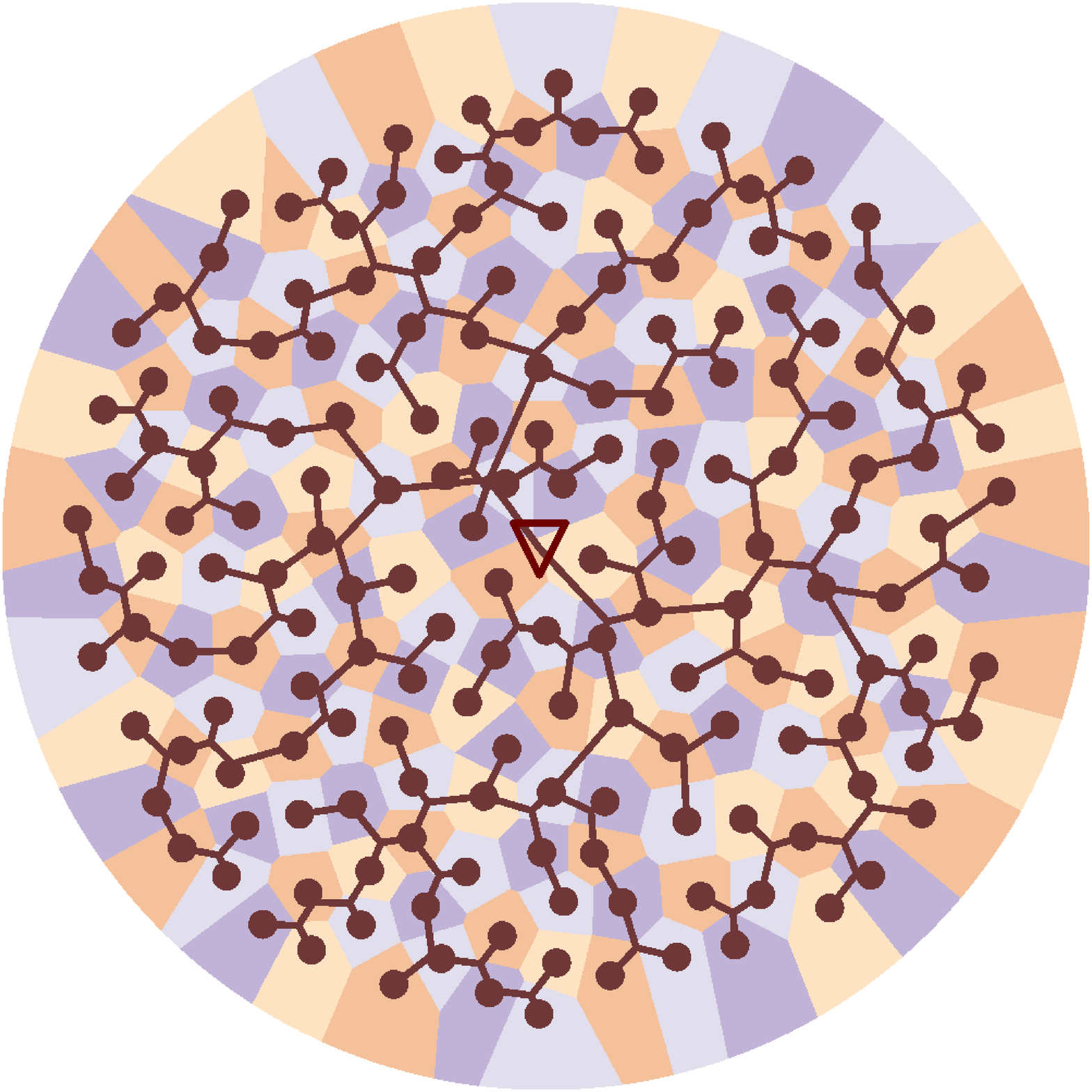}
	\label{p1_c100}
}
&
\subfloat[]{
	\includegraphics[width=3cm]{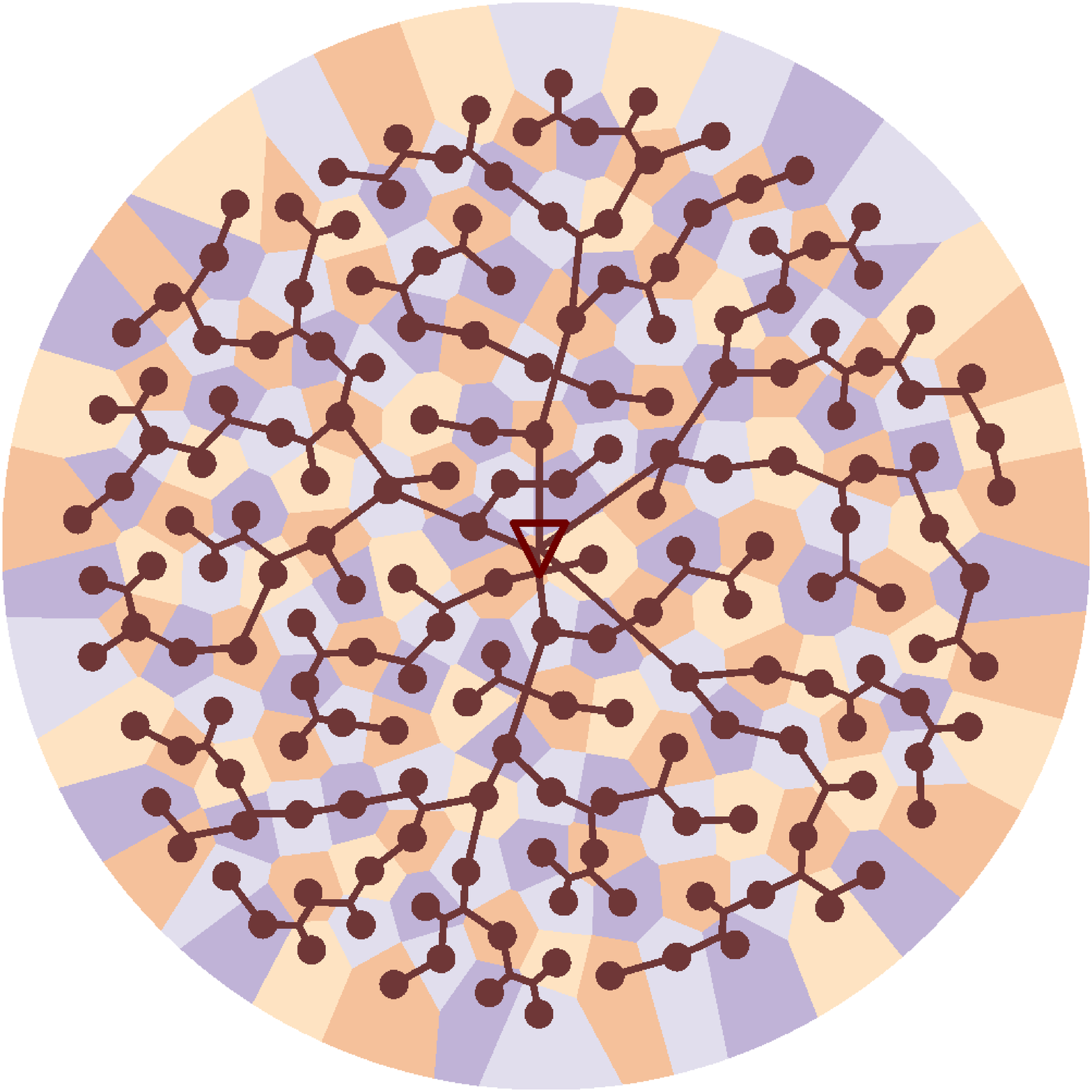}
	\label{p1_c190}
}
\\
\parbox{1.9cm}{\centering{$U_0 = 0.6$}}
&
\subfloat[]{
	\includegraphics[width=3cm]{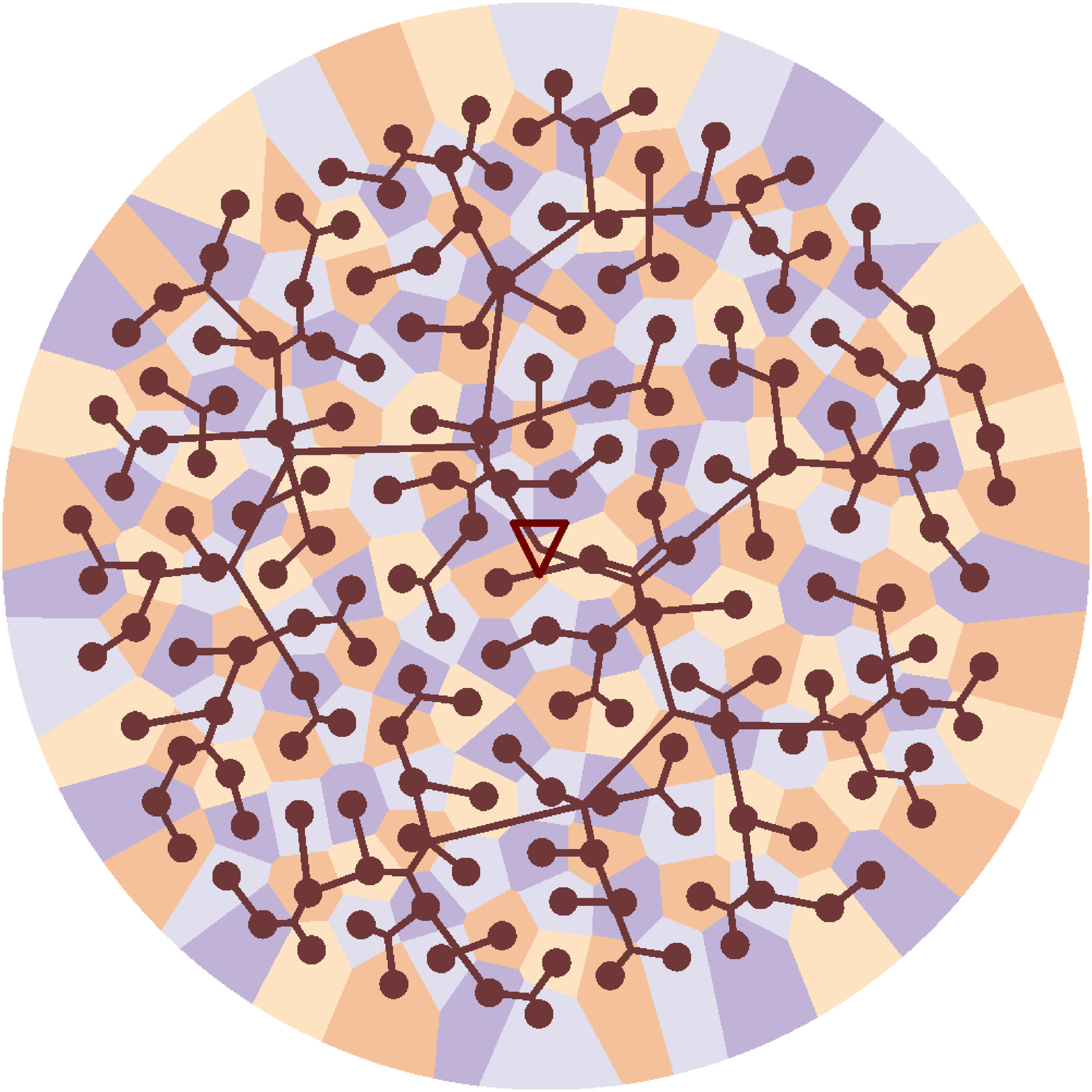}
	\label{p4_c100}
}
&
\subfloat[]{
	\includegraphics[width=3cm]{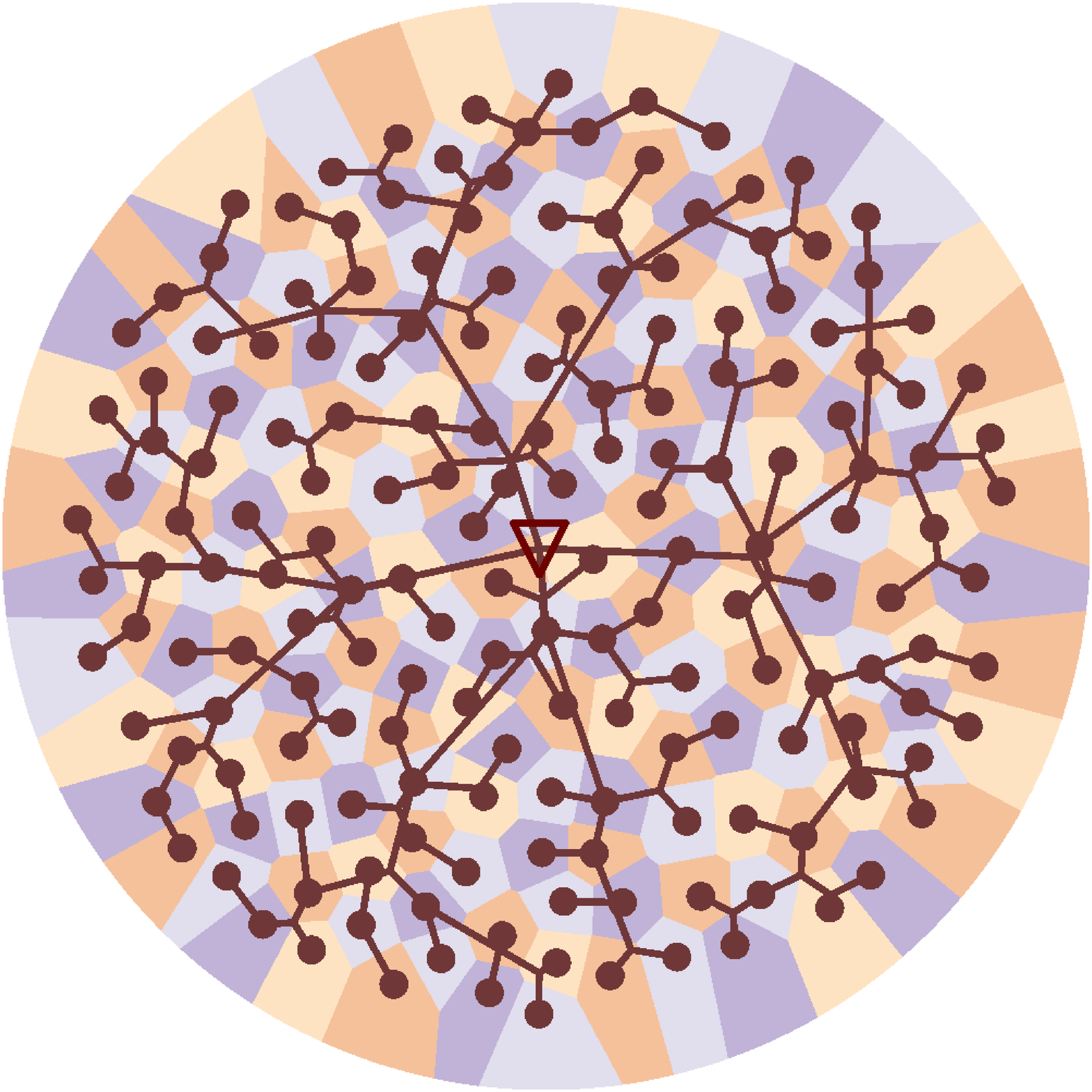}
	\label{p4_c190}
}
\end{tabular}
\caption{
	(Color)
	Optimal configurations for several constraints on hierarchical balance and two optimization weights for fitness $F$.
	The BHC seed is the same as in Fig. \ref{large_optimal_networks_circ}.
	We choose the weights ${(C_L,~C_H)~=~(1, 9)}$ so that the contribution from $H$ is not dominated by the contribution from $L$.
}
\label{large_constrained_networks}
\end{figure}

\begin{figure*}[ht]
\centering
\vspace*{0.5cm}
\begin{tabular}{c@{\hskip 1cm}c}
\parbox{4cm}{
	\centering{
		$\lambda_L$ for Total Network Length $L$ ${[F(1,~0)]}$
	}
}
\vspace*{0.9cm}
&
\parbox{4cm}{
	\centering{
		$\lambda_L$ for Total Network Length $L$ and Average Path Length $H$ ${[F(1,~9)]}$
	}
}
\\
\subfloat[]{
	\includegraphics[width=7.8cm]{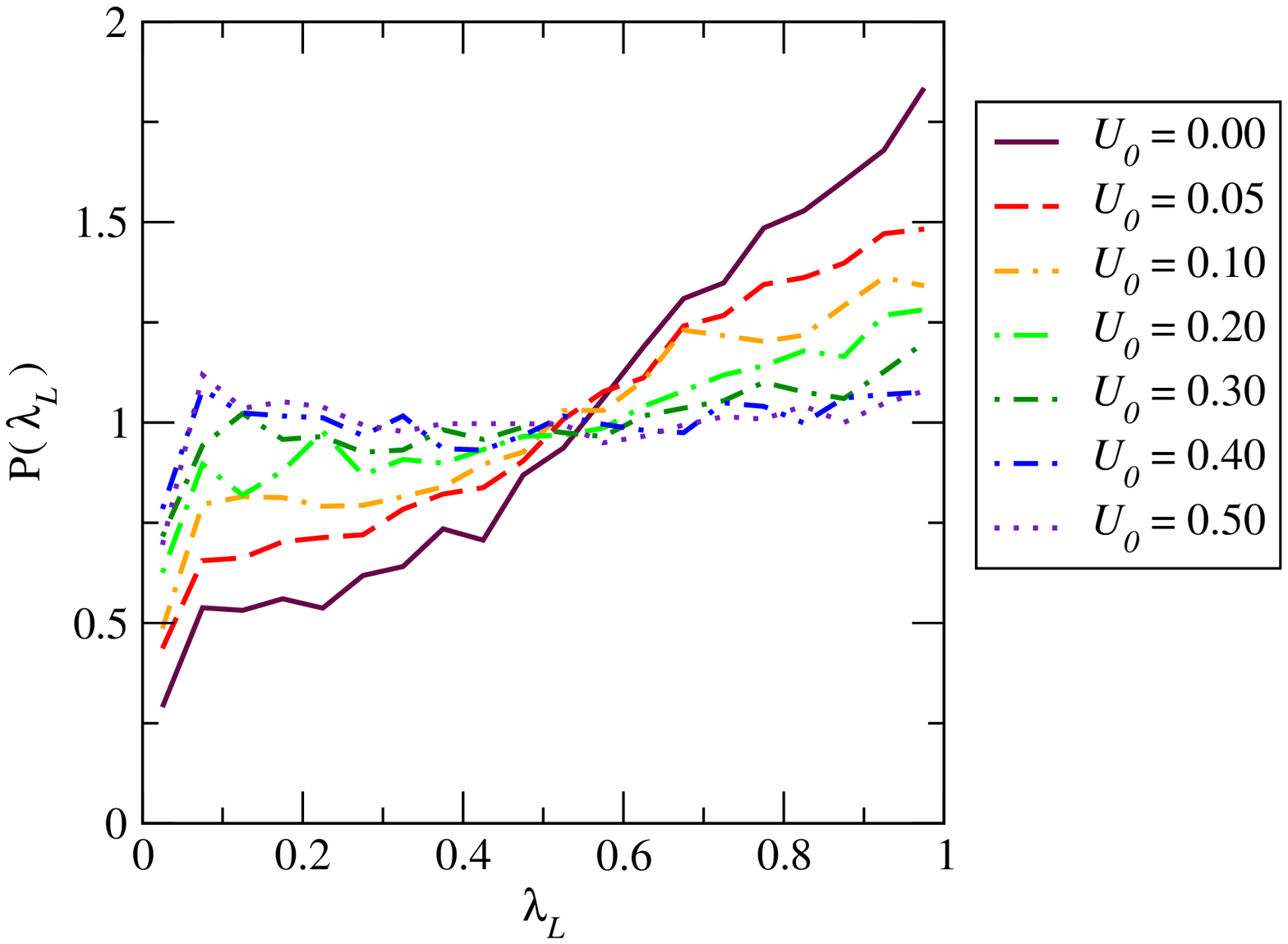}
	\label{constrained_lambdaL_length}
}
&
\hspace{1cm}
\subfloat[]{
	\includegraphics[width=7.8cm]{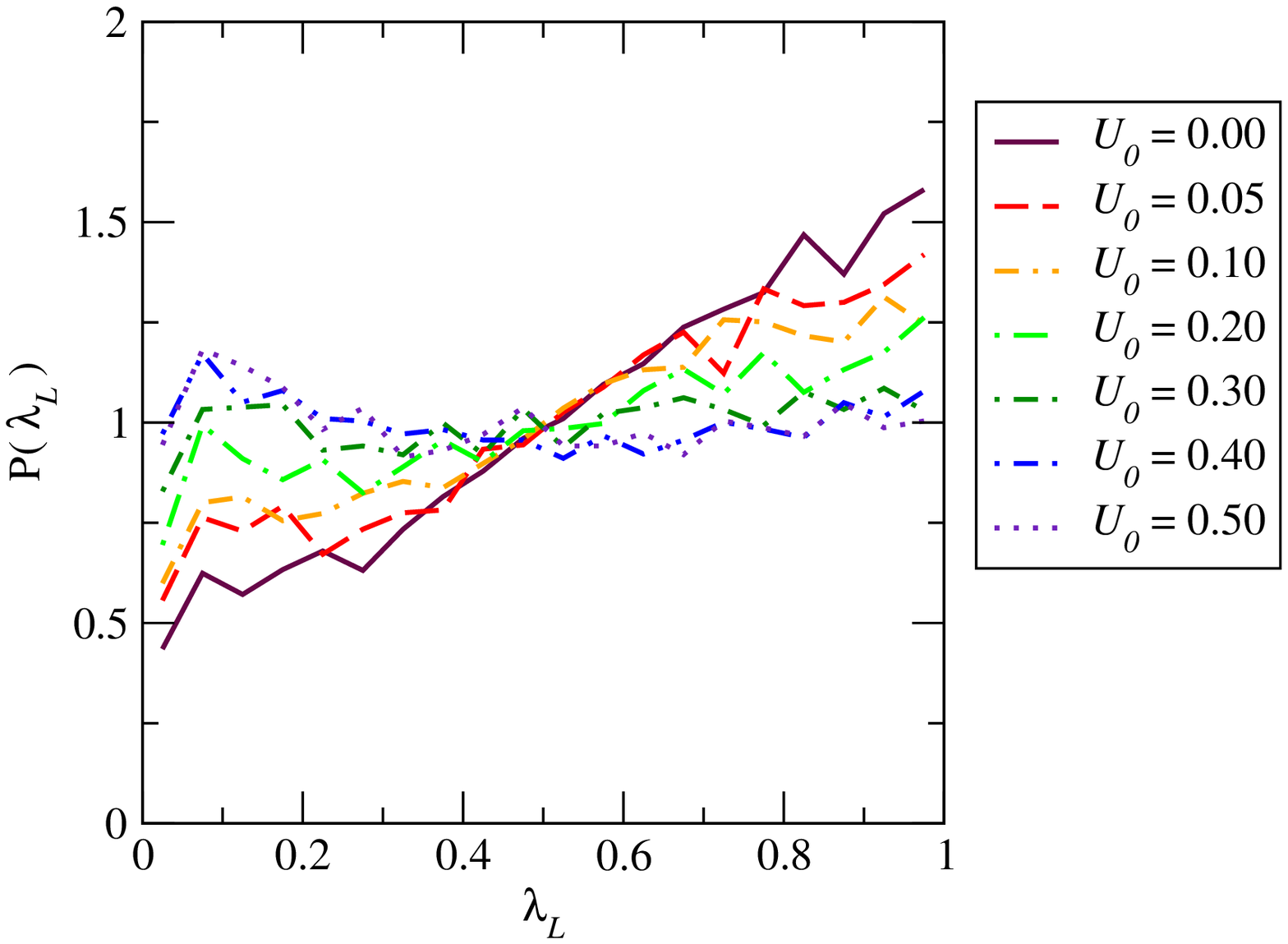}
	\label{constrained_lambdaL_path}
}
\end{tabular}
\caption{
	(Color)
	Distributions for several thresholds of hierarchical unbalance $U_0$.
	(a)~Distributions of $\lambda_L$ optimized solely for total network length $L$ ${[F(1, 0)]}$.
	(b)~Distributions for $\lambda_L$ optimized for total network length $L$ and average path lengths $H$ ${[F(1, 9)]}$.
	All plots are averaged over 200 realizations of service volume distributions.
}
\label{constrained_lambdaL}
\end{figure*}

Imposing a constraint on hierarchical balance leads to configurations that reflect more realistic asymmetry in branching lengths.
Hierarchical balance, which equalizes the number of service volumes that each sibling segment supplies, is related to the blood flow that is required to deliver resources and effectively limits the asymmetry of sibling radii.
In Fig. \ref{large_constrained_networks} we show results for several thresholds for the constraint on hierarchical balance.
Decreasing the threshold $U_0$  yields more realistic distributions for $\lambda_L$, as shown in Fig. \ref{constrained_lambdaL}.
Because these networks are embedded in {2-$D$}, decreasing $U_0$ can also result in more crossings between segments at different levels.
By comparing Figs. \ref{constrained_lambdaL_length} and \ref{constrained_lambdaL_path}, we see that the constraint on hierarchical balance leads to similar results independent of the weight of average path length $H$ in configuration fitness.
Instead of contributing significantly to fitness, $H$ is effectively optimized through hierarchical balance.

\begin{figure}[ht]
\centering
\vspace*{1cm}
	\includegraphics[width=7.3cm]{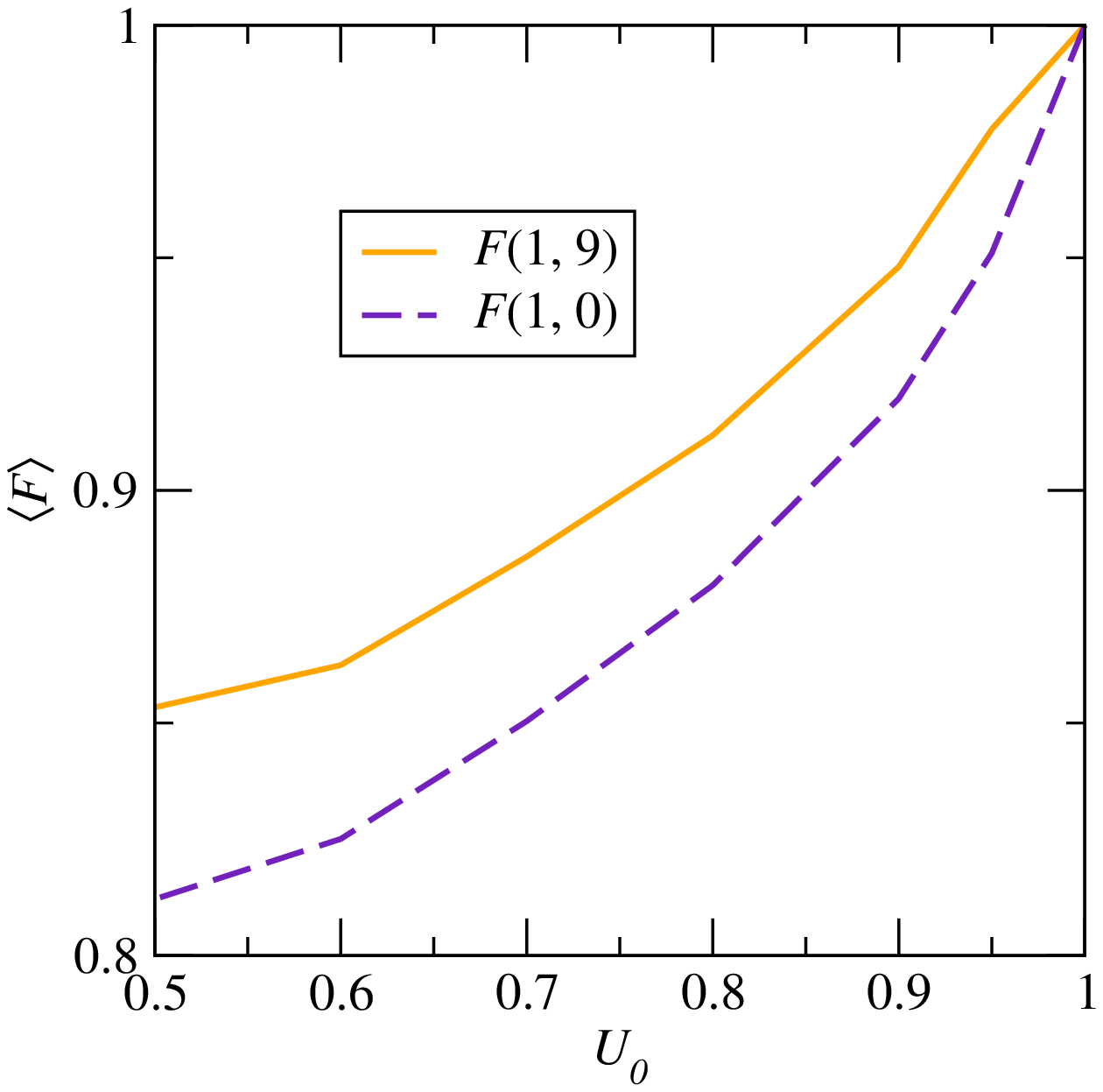}
\caption{
	(Color)
	Average fitnesses $F$ for ensembles of 100 optimized configurations.
	Configurations are more fit if a greater hierarchical unbalance $U_0$ is allowed.
}
\label{hierarchicalFitness}
\end{figure}

While enforcing hierarchical balance leads to more realistic branching and length asymmetry distributions, it is not necessary to have a maximally balanced hierarchy.
In Fig.\ref{hierarchicalFitness}, we show that lowering the threshold $U_0$ reduces network fitness.
As the constraint on hierarchical balance $U_0$ decreases, the average fitness also decreases.
Since the distribution of $\lambda_L$ is approximately uniform around $U_0 \approx 0.7$ and below, the best value for the hierarchical constraint is ${U_0~\approx~0.7}$ because this yields the fittest networks that have uniform distributions for length asymmetry.

\section{Discussion}
\label{discussion_section}

With our determination of branch point positions and exploration of distinct hierarchical configurations, we can remark on several consequences that follow from the general properties of optimized networks.
Organizing the lengths between branch points to fill {2-} or {3-$D$} space with capillaries inevitably leads to asymmetries and unbalanced networks \cite{banavar10}.
Strictly symmetric and balanced networks are either inefficient in materials or not space-filling.
For example, in the H-tree all children branch orthogonally from the parent, resulting in inefficient paths.
Other networks with more efficient paths lead to capillaries that are equidistant from the source, which could cover the surface of a sphere but not fill its volume.
For the optimal, space-filling networks that we explore, we impose a constraint that pushes the network toward hierarchically balanced branching structures but does not require maximum balance.
One can imagine other interesting metrics for hierarchical balance, but we concentrate on how a maximum degree of unbalance $U_0$ affects the structure of the network.
This guarantees a minimum level of balance in the hierarchy but still allows freedom in the search for optimal networks, as well as nonuniformity in the hierarchical balance.

We construct a seed configuration that builds a network to ensure maximal hierarchical balance while maintaining efficient contiguity of subtrees.
Configurations that tend to be hierarchically balanced, such as the BHC configuration (where the constraint is implicit in the construction algorithm) or optimized configurations that limit unbalance, do not show a strong skew toward symmetric branching in lengths.
This hierarchical balance may result from gradual, incremental growth as an individual organism matures and ages.
Nearby vessels grow to supply resources to new tissue, resulting in contiguous subtrees and favoring routes that reduce path lengths and avoid a single, meandering artery that branches directly to capillaries.

Other computational models approach the growth and optimization of space-filling networks in different ways.
Although there are many algorithms to generate structure that do not intentionally optimize network architecture or space-filling properties, near-optimal configurations may emerge spontaneously from certain simple rules.
Examples of such pattern formation processes and associated algorithmic rules include models for both angiogenesis \cite{Meinhardt76, Yao07}, as well as vasculogenesis (in terms of chemotaxic \cite{Serini03, Gamba03}, mechanical substratum \cite{Manoussaki96}, and cellular Potts models \cite{Merks06, Merks09}).
However, these models do not adequately address our focus on branching length asymmetries for efficient, hierarchical, space-filling networks.
Specifically, the pattern formation model for angiogenesis does not incorporate consistent space-filling service volumes, only space-filling arterial structure.
The arterial structure fills some regions so that they are devoid of capillaries, while multiple tips converge to the same location elsewhere.
The models for vasculogenesis do not optimize the development of a hierarchical branching network.
However, dynamic vascular remodeling \cite{HuCai13} can form structures both with and without closed loops while maintaining a uniform distribution of capillaries, although the optimal structures also suffer from large arteries branching directly to capillaries.
We extend these models to understand the asymmetric lengths of adjacent segments in vascular networks and how these relate to space-filling service volumes.

Because of the many different factors and interactions that influence the structure of the cardiovascular system, our basic model can be expanded in many directions.
Radius information can be incorporated into optimized networks by requiring flow to be uniform in all terminal service volumes.
By including radius information, blood flow as well as more appropriate structural and energetic costs can lead to revised optimization principles, which require the calculation of the \emph{weighted} Fermat point (e.g., see \cite{Shen08}) and has been explored previously in a limited, local context \cite{Zamir78, Zamir00, Zamir05}.
Note that lowering the threshold $U_0$ tends to increase the minimum number of branching levels between the heart and capillaries.
Less drastic hierarchical unbalance implies that the ratios of parent-child radii $\beta = r_c/r_p$ should be near 1 (symmetric branching).
This translates the global, topological property into a local branching quantity.

We do not expect that increasing the dimensionality of our networks to {3-$D$} would change the qualitative results for branching asymmetry in length ($\lambda_L$) with hierarchical balance (specifically $U_0$).
However, the numerical location for an optimal trade-off between fitness and balance may shift.
Studies of large vessels (near the heart) show these vessels to be planar \cite{Wischgoll09}), but the planarity cannot always hold across the entire network if tips must fill a {3-$D$} space.
Still, in the absence of obstacles, all optimization conditions enforce planarity in {3-$D$} for branch points in their local context.
Introducing regions where the network is prohibited (e.g. through bones, organs or from self-avoidance) constrains the Fermat point to the surface of a sphere or some other shape \cite{Zachos14sphere, Zachos14flat}.

While the topological change of allowing loops introduces many complications to the properties of flow and hierarchical labels \cite{Mileyko12}, such a modification can be beneficial in understanding reticulated vascular structures.
Loops are especially important when considering network robustness (i.e., resilience to damage) within organs and leaves \cite{Corson10, Katifori10}) or pathological growth in tumors \cite{Herman11, Savage13}.
These types of network properties can be included in future models.

Locally, the position of a branching junction minimizes the sum of vessel lengths in our model.
Globally, we impose a threshold on the minimum hierarchical balance, which reduces the differential blood flow into sibling segments.
Although real vascular networks consist almost entirely of bifurcations (although there is rapid, asymmetric branching from the aorta to capillaries through coronary arteries), the iterative approach described in Sec. \ref{local_optimization_of_positions_subsection} can lead to low numbers of bifurcating junctions for some candidate networks.

Limiting the degree of unbalance in the hierarchy does not continue to shift the distribution of $\lambda_L$ away from symmetry (${\lambda_L~\approx~1}$) below ${U_0~\approx~0.7}$, which suggests that there is an appropriate trade-off between the hierarchical balance threshold $U_0$ and configuration fitness $F$ that does not require perfect symmetry for an efficient network structure.
The increased cost of the network in Fig. \ref{hierarchicalFitness} is similar for both curves, implying that the increase mostly comes from total network length $L$.

The large number of distinct bifurcating hierarchies necessitates that we carefully choose and execute the algorithms for searching the space of possible configurations.
Consequently, we construct a favorable starting point and concentrate computational resources on regions that are most likely to contain optimal configurations.
Using the numerical implementations in sections \ref{model_section} and \ref{configuration_section}, we identify optimal networks and study the length properties of individual segments within the context of a network with space-filling terminal service volumes.

Our results have many implications for how vascular networks fill space efficiently.
We exhaustively explore fitness landscapes for small networks and carefully guide the sampling of the space of hierarchies for large networks in order to determine near-optimal configurations.
Our results show that strict hierarchical balance is not optimal for the architecture of cardiovascular networks.
Furthermore, there is a trade-off between hierarchical balance (which is related to symmetric branching in radius at the local level) and the distribution for branching in lengths that shows the connection between the space-filling and efficiency requirements of the network.
By incorporating radius and flow information, as well as growth patterns that incorporate obstacles and loops, we can continue to build on present models to better understand vascular architecture and gain insights for its effects on resource delivery, metabolic scaling, aging, and repair after damage.

\appendix
\section{Similarity measure to compare hierarchical groupings between configurations}

While collapsing the landscape of measures to a single dimension informs us about the typical distribution of configurations, it retains no information about the relation of the hierarchies between different trees.
To address this issue, we define a measure of similarity to compare how two hierarchies group the same set of tips.
This measure is normalized such that similar hierarchies and groupings of service volumes have a similarity score near 1, while hierarchies that group service volumes in very different ways have a similarity score near 0.
To meet these guidelines, we perform a simple count of the number of identical subtree groupings between two hierarchies and normalize by the maximum possible number that could be shared if the trees were identical.
In accordance with these properties, define the similarity $S(A, B)$ between two configurations $A$ and $B$ as
\begin{equation*}
S(A, B) \equiv \frac{\sigma(A, B)}{\max\left\{\sigma(A, A), \sigma(B, B)\right\}}
\end{equation*}
\begin{equation*}
\sigma(X, Y) = \sum_{\substack{\rm{subtree~}m \\ \rm{in~network~}X}}~~\sum_{\substack{\rm{subtree~}n \\ \rm{in~network~}Y}}\left(I_{m \subseteq n} + I_{n \subseteq m}\right)
\end{equation*}
where $I_s$ is the indicator function (1 if statement $s$ is true and 0 otherwise) and \textit{subtree} refers to the set of tips in that particular subtree.

\begin{figure*}[ht]
\centering
\vspace*{1cm}
\subfloat[]{
	\includegraphics[width=5.6cm]{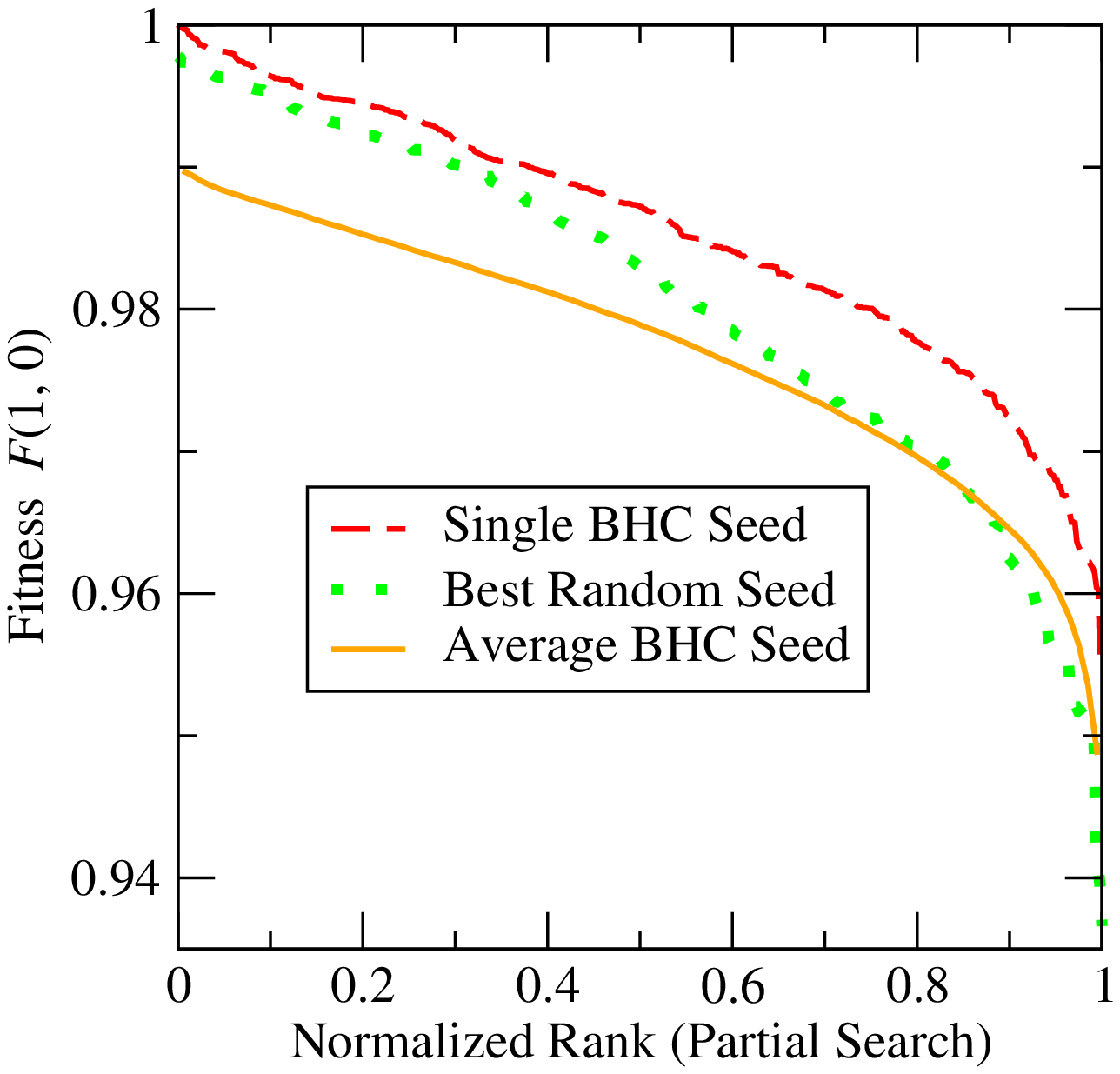}
	\label{guess_and_random_runs}
}
\hspace{1.6cm}
\subfloat[]{
	\includegraphics[width=8.6cm]{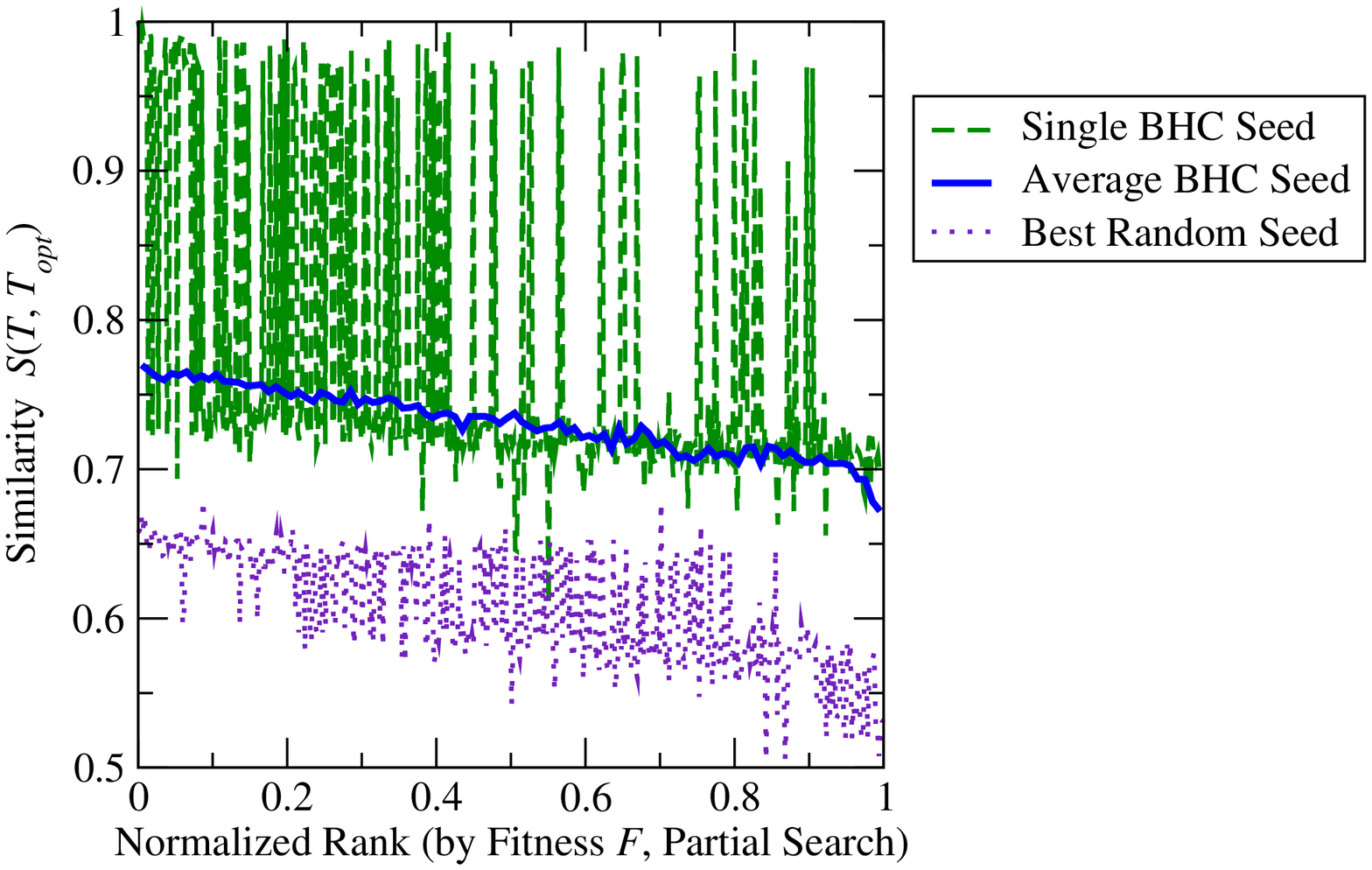}
	\label{large_ave_meas_sim}
}
\caption{
	(Color)
	Partial search trajectories in the the space of hierarchies.
	(a) Ranked fitness landscapes  for the partially explored hierarchical space during optimization for total network length $L$ ${[F(1,~0)]}$.
	(b) Corresponding similarity measures.
	All quantities are calculated with respect to the most optimal configuration found over all runs.
	The solid line shows the average fitness over 100 trajectories from the BHC for a system with 36 service volumes.
	Individual paths are for a network of 48 service volumes.
	One path begins with the BHC configuration and the other begins with a random seed (the best performing of 5 random seeds).
}
\label{similarity_fig}
\end{figure*}

Configurations that have a worse fitness measure are less similar to the optimal configuration, as shown in Fig. \ref{similarity_fig}.
However, note that similarity $S(T_i, T_{opt})$ is not a monotonic function when rank $i$ is defined by the configuration's measure.
For example, consider hierarchy A, which may be very similar to a hierarchy B, which itself is very similar to C.
Then it is possible that A and C are less similar to each other than each is to B, yet both are ranked higher than B with respect to a particular measure.
This also means that optimal configurations are not always a single swap or regraft away from all near-optimal configurations, i.e. local minima are possible.
The average similarity in Fig. \ref{large_ave_meas_sim} does not approach 1, indicating that the subtree grouping of service volumes can be very different between networks that are nearly optimal (${F \approx F_{opt}}$).
Some of the stratification into distinct levels of similarity is apparent in Fig. \ref{guess_and_random_runs} for the single trajectories (the ``Single BHC Seed'' and ``Best Random Seed'').

\section{Network size and shape}
\label{size_and_shape_section}

\begin{figure*}[ht]
\centering
\vspace*{1.1cm}
\subfloat[]{
	\includegraphics[width=7cm]{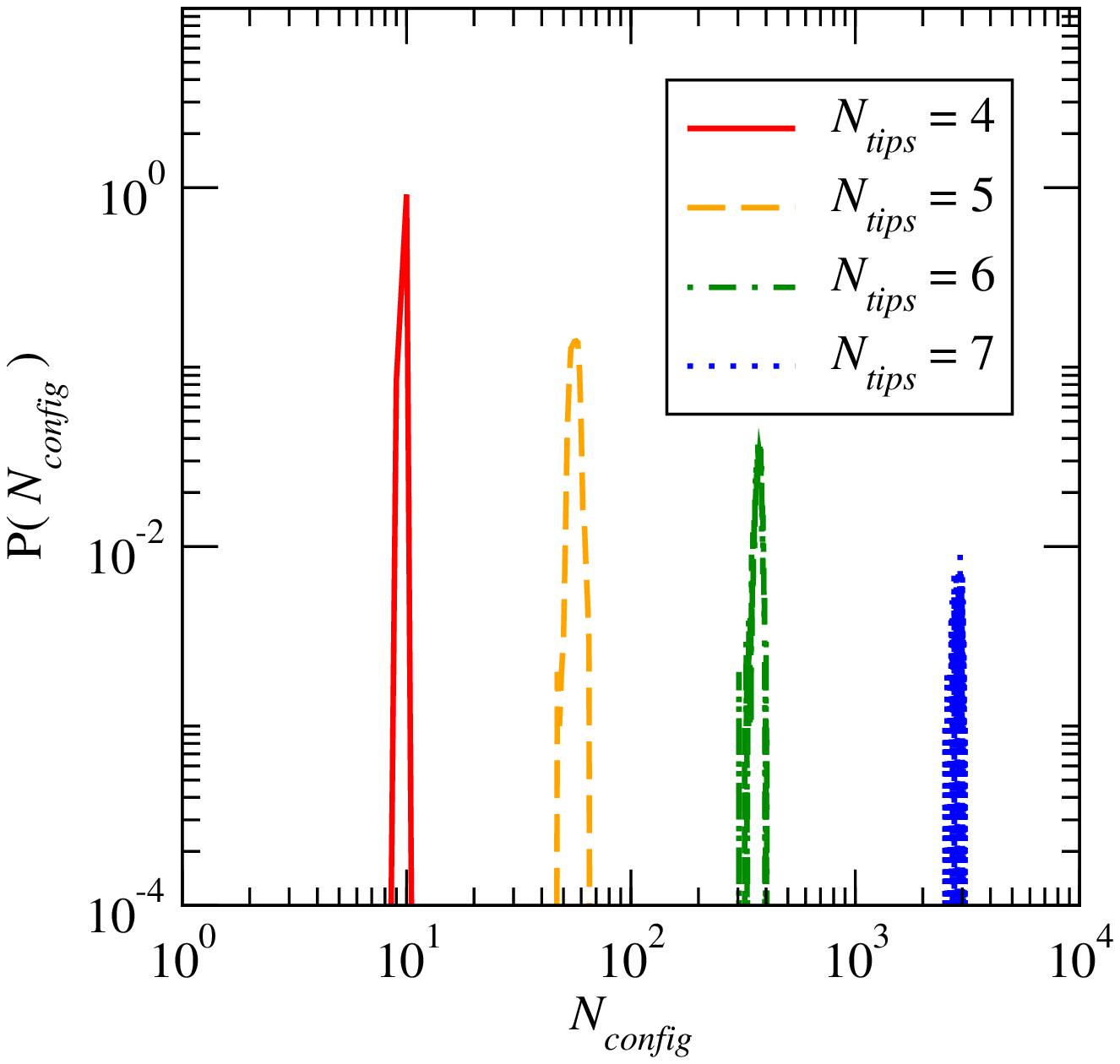}
	\label{unique_trees_small}
}
\hspace{1.5cm}
\subfloat[]{
	\includegraphics[width=7.2cm]{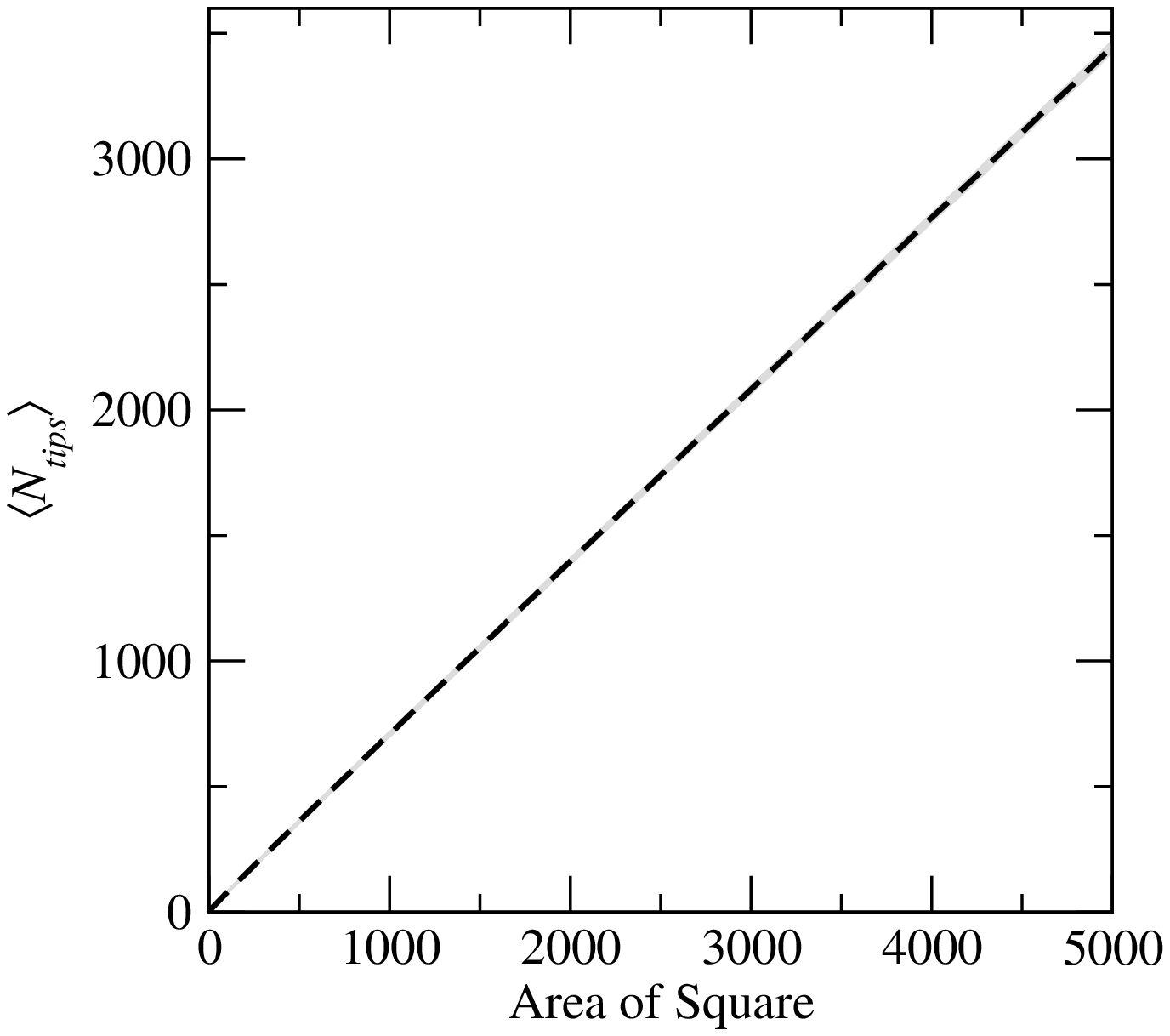}
	\label{tips_per_area}
}
\caption{
	(Color)
	Network properties with increased size.
	(a)~The number of unique configurations $N_{config}$ for each fixed number of service volumes is narrowly distributed relative to the increased number of configurations from introducing an additional service volume.
	(b)~The average number of service volumes increases directly proportional to the total area.
The solid grey line represents the standard deviation about the mean.
}
\end{figure*}

\begin{figure}[ht]
\vspace*{1cm}
\centering
\includegraphics[width=8.3cm]{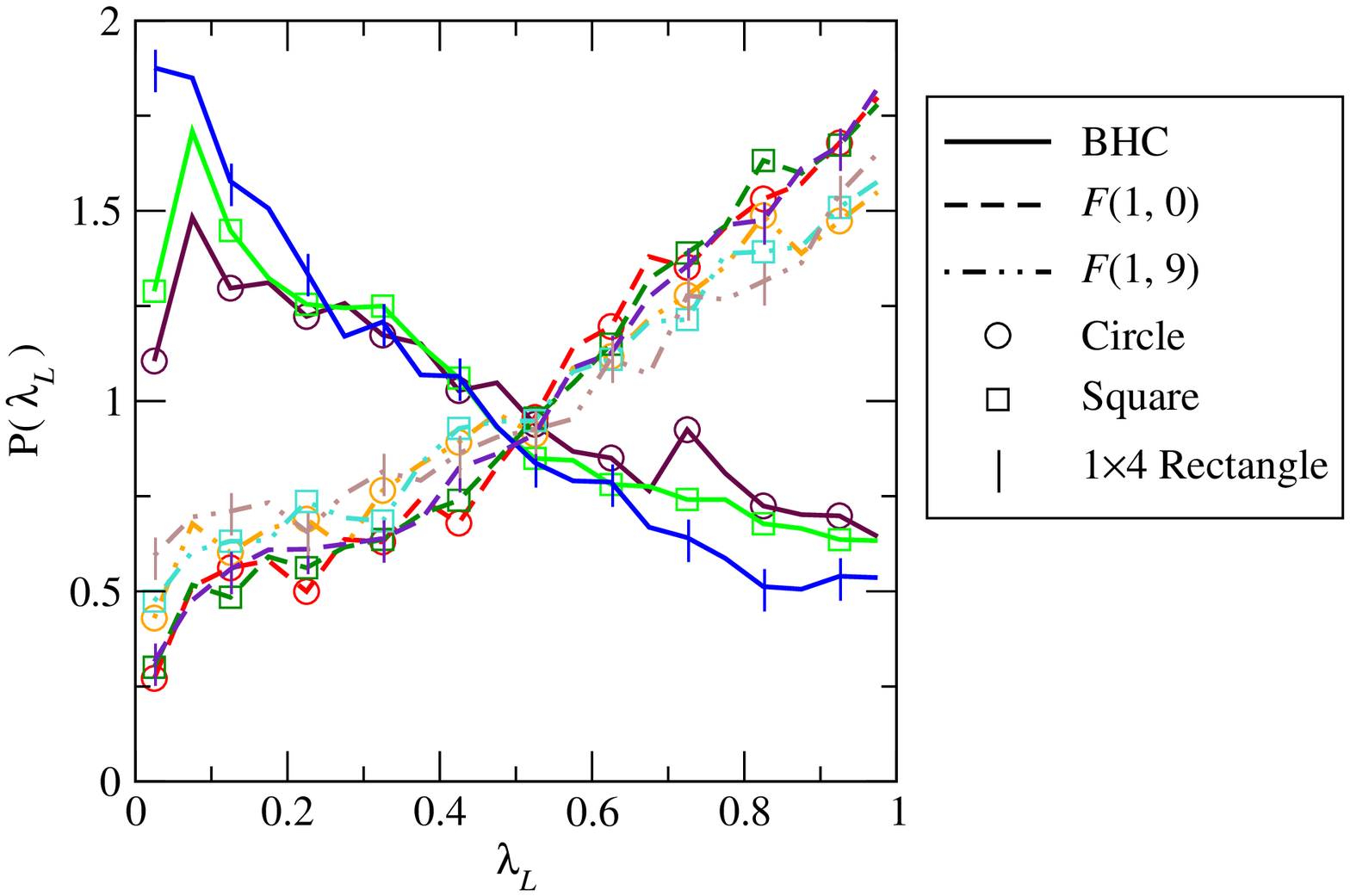}
\label{no_constraint_shapes_optimal_lambda_L}
\caption{
	(Color)
	The distributions for $\lambda_L$ in optimized networks [${F(1,~0)}$ and ${F(1,~9)}$] bounded in circular, square, and ${1 \times 4}$ rectangular areas are all similarly skewed \emph{toward} symmetry.
	However, the distributions in BHC networks are all skewed \emph{away} from symmetry.
}
\label{no_constraint_optimal_lambda_L_sizes}
\end{figure}

In Fig. \ref{unique_trees_small} we show the distribution of the number of distinct hierarchies after consolidating degenerate bifurcations for these ensembles of fixed numbers of service volumes.
The variance in the number of unique configurations increases with network size, but a dominating contribution that increases the number of configurations comes from adding a service volume.
Not surprisingly, the average total number of service volumes $\langle N_{tips}\rangle$ scales linearly with the total area (see Fig. \ref{tips_per_area}).
\begin{figure*}[ht]
\captionsetup[subfigure]{labelformat=empty}
\vspace*{0.5cm}
\centering
\begin{tabular}{ccc}
\parbox{4cm}{
	\centering{BHC Seed}
}
&
\parbox{4cm}{
	\centering{Total Network Length $L$ $[{F(1,~0)]}$}
}
&
\parbox{4cm}{
	\centering{Total Network Length $L$ and Mean Path Length $H$ ${[F(1,~9)]}$}
}
\\
\subfloat[]{
	\includegraphics[width=3cm]{singleNetOptRec_c1.0-0.0-0.0_hierSym0.000_side10.1circ_00001_guessSym.eps}
	\label{circ_nsc_again}
}
&
\subfloat[]{
	\includegraphics[width=3cm]{singleNetOptRec_c1.0-0.0-0.0_hierSym0.000_side10.1circ_00001_optGraft.eps}
	\label{circ_c100_again}
}
&
\subfloat[]{
	\includegraphics[width=3cm]{singleNetOptRec_c1.0-9.0-0.0_hierSym0.000_side10.1circ_00004_optGraft.eps}
	\label{circ_c190_again}
}
\\
\subfloat[]{
	\includegraphics[width=3cm]{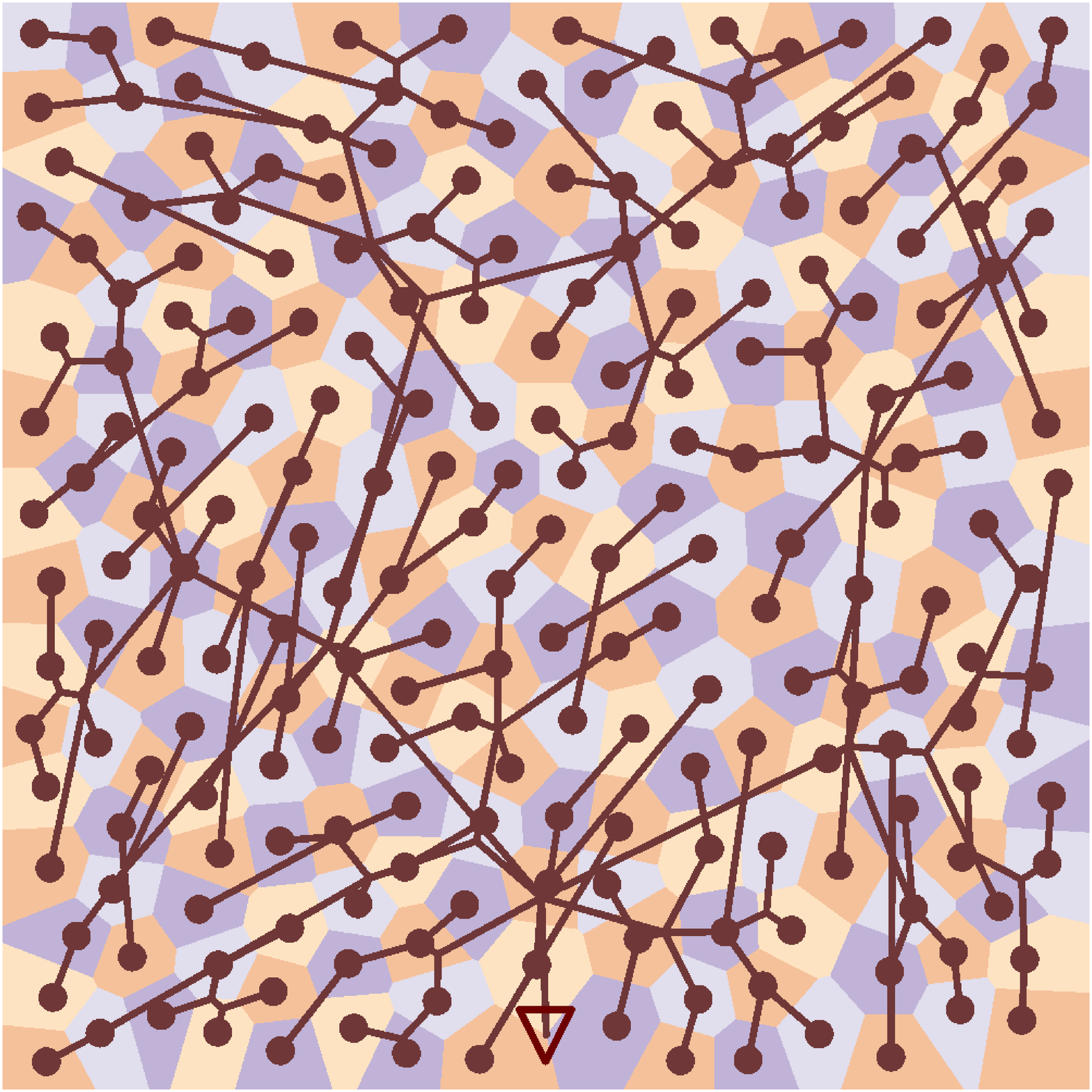}
	\label{square_nsc}
}
&
\subfloat[]{
	\includegraphics[width=3cm]{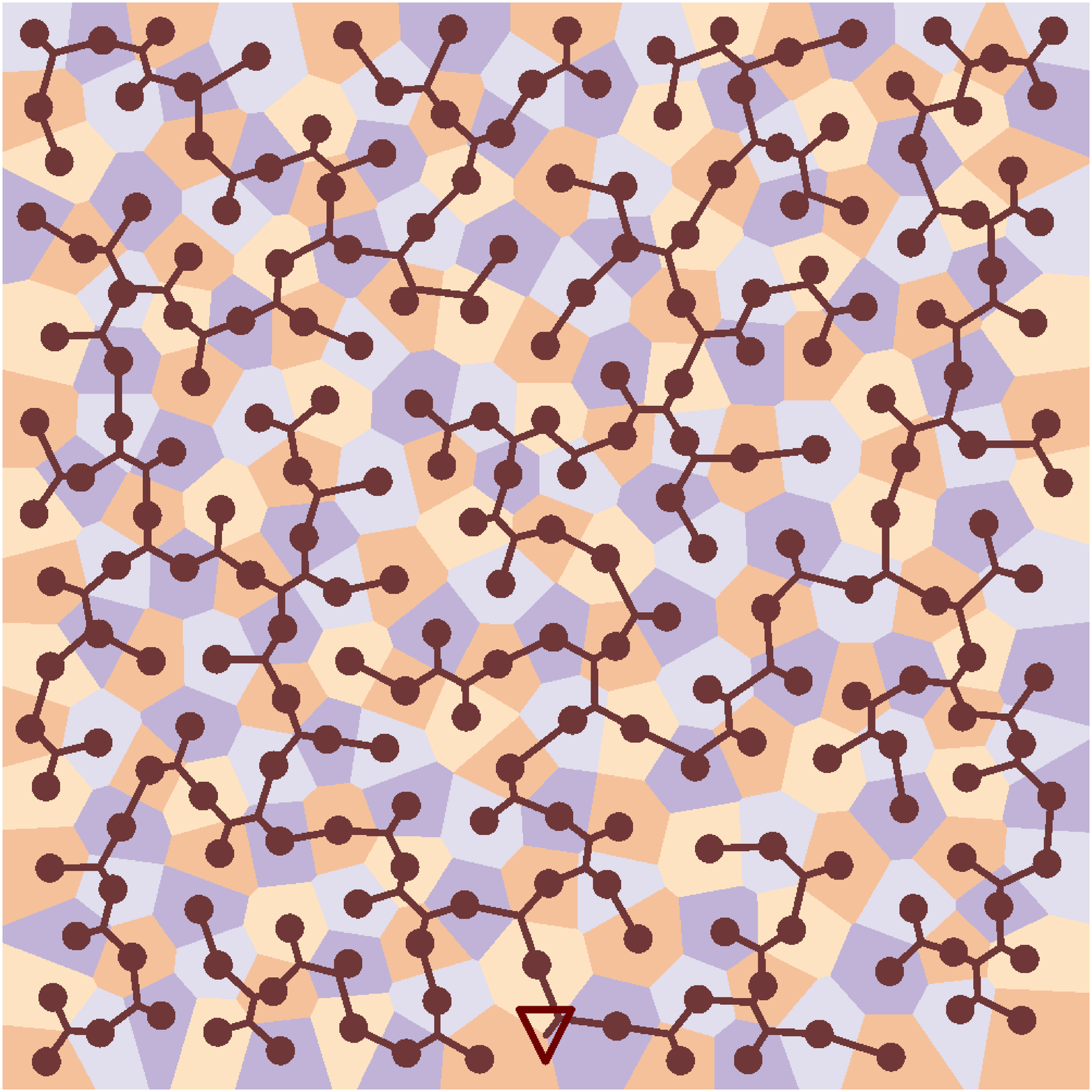}
	\label{square_c100}
}
&
\subfloat[]{
	\includegraphics[width=3cm]{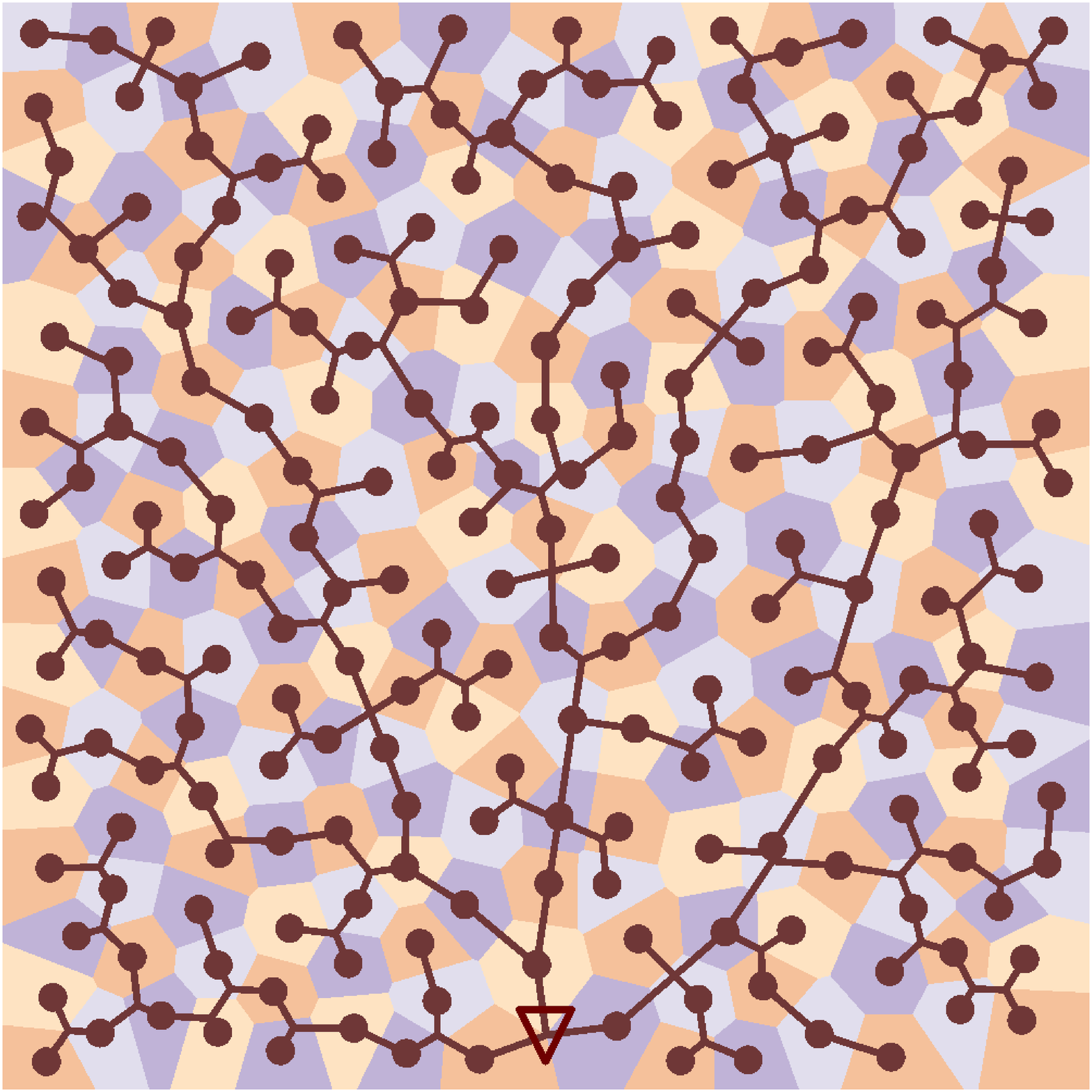}
	\label{square_c190}
}
\\
\subfloat[]{
	\includegraphics[width=3cm]{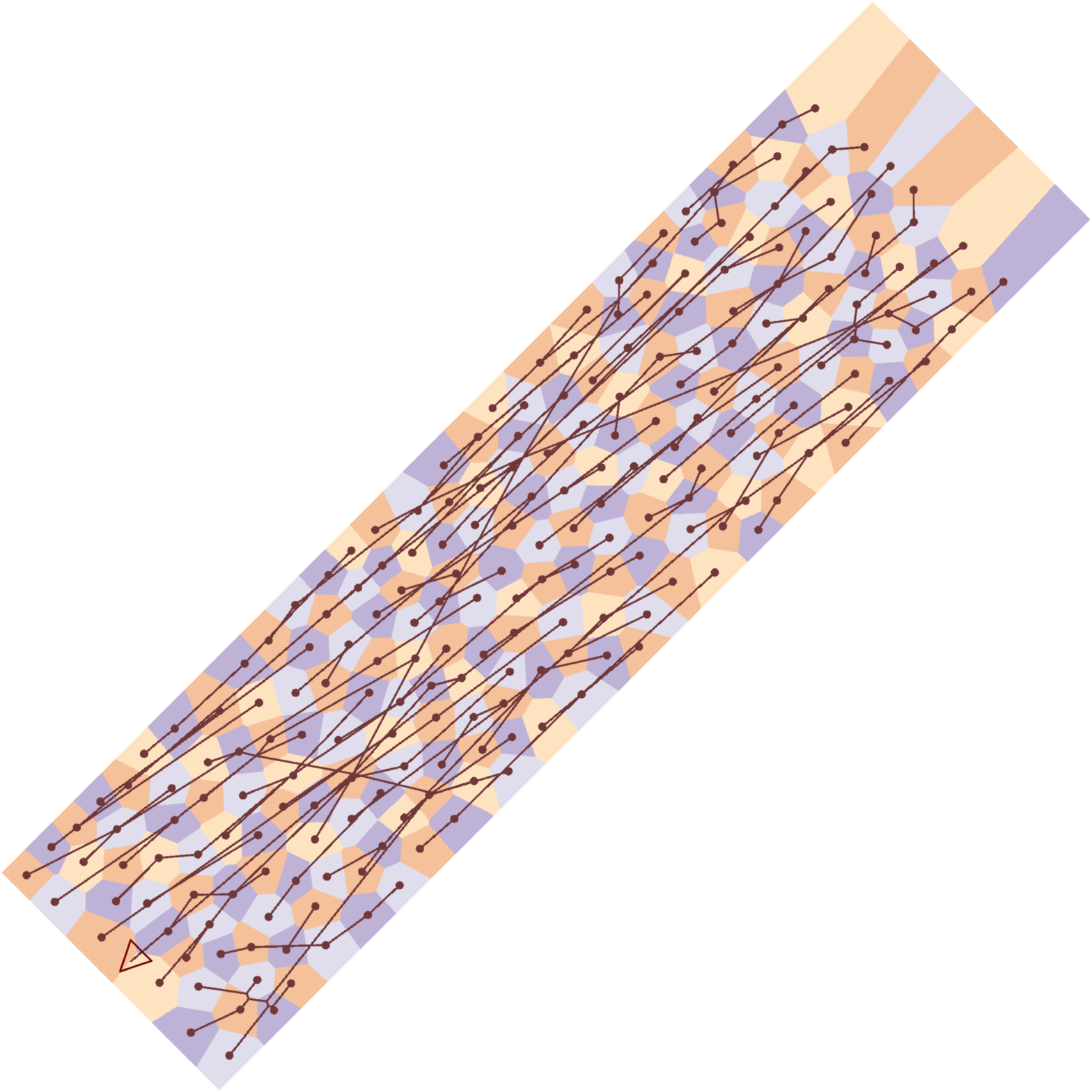}
	\label{oneFour_nsc}
}
&
\subfloat[]{
	\includegraphics[width=3cm]{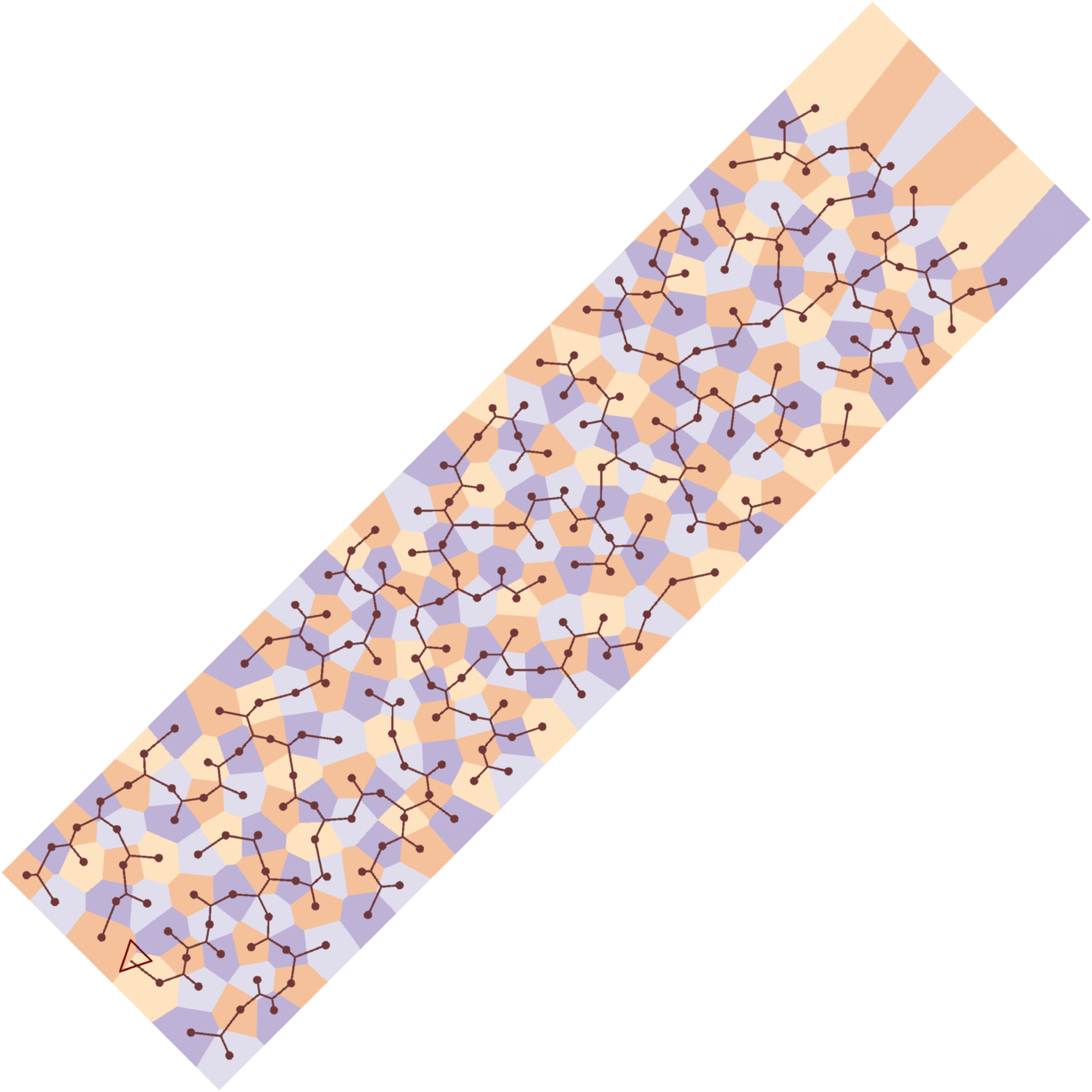}
	\label{oneFour_c100}
}
&
\subfloat[]{
	\includegraphics[width=3cm]{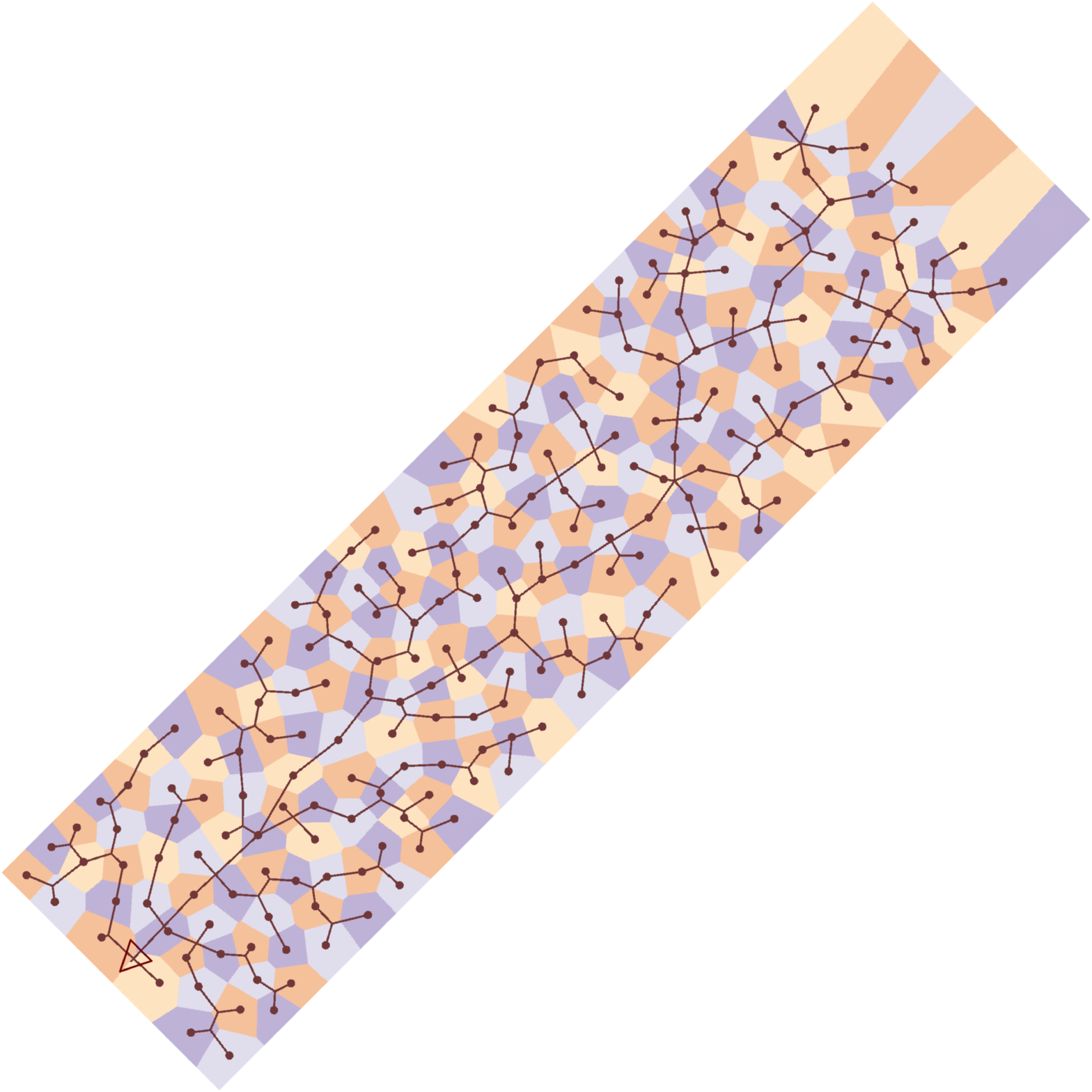}
	\label{oneFour_c190}
}
\end{tabular}
\caption{
	(Color)
	Optimal configurations for several shapes and two fitness measures.
	Both isotropic (circular) and anisotropic (rectangular) areas exhibit similar space-filling strategies for near-optimal configurations.
	We choose the weights ${(C_L,~C_H)~=~(1, 9)}$ so that the contribution from $H$ is not dominated by the contribution from $L$.
}
\label{large_optimal_network_shapes}
\end{figure*}
In Fig. \ref{no_constraint_optimal_lambda_L_sizes} we show distributions for branching length asymmetry $\lambda_L$.
All distributions for optimized networks that have no constraint on the hierarchy (i.e. $U_0 = 1$) exhibit a strong skew toward symmetry.

\section{Parent-child length ratio $\gamma$}
\label{gamma_section}

\begin{figure*}[ht]
\centering
\vspace*{1cm}
\subfloat[]{
	\includegraphics[width=6cm]{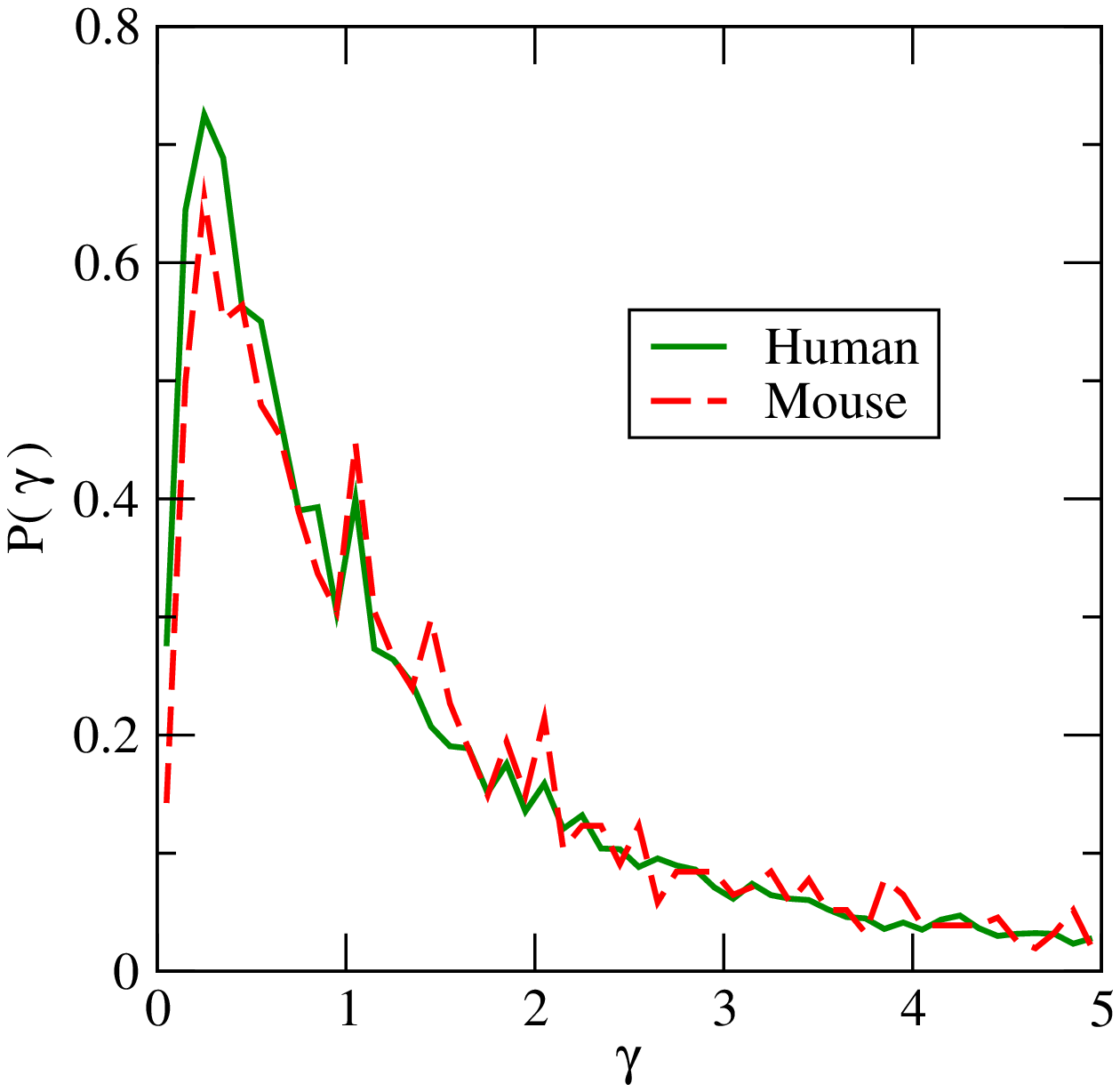}
	\label{gamma_dicom}
}
\hspace{1.6cm}
\subfloat[]{
	\includegraphics[width=8.2cm]{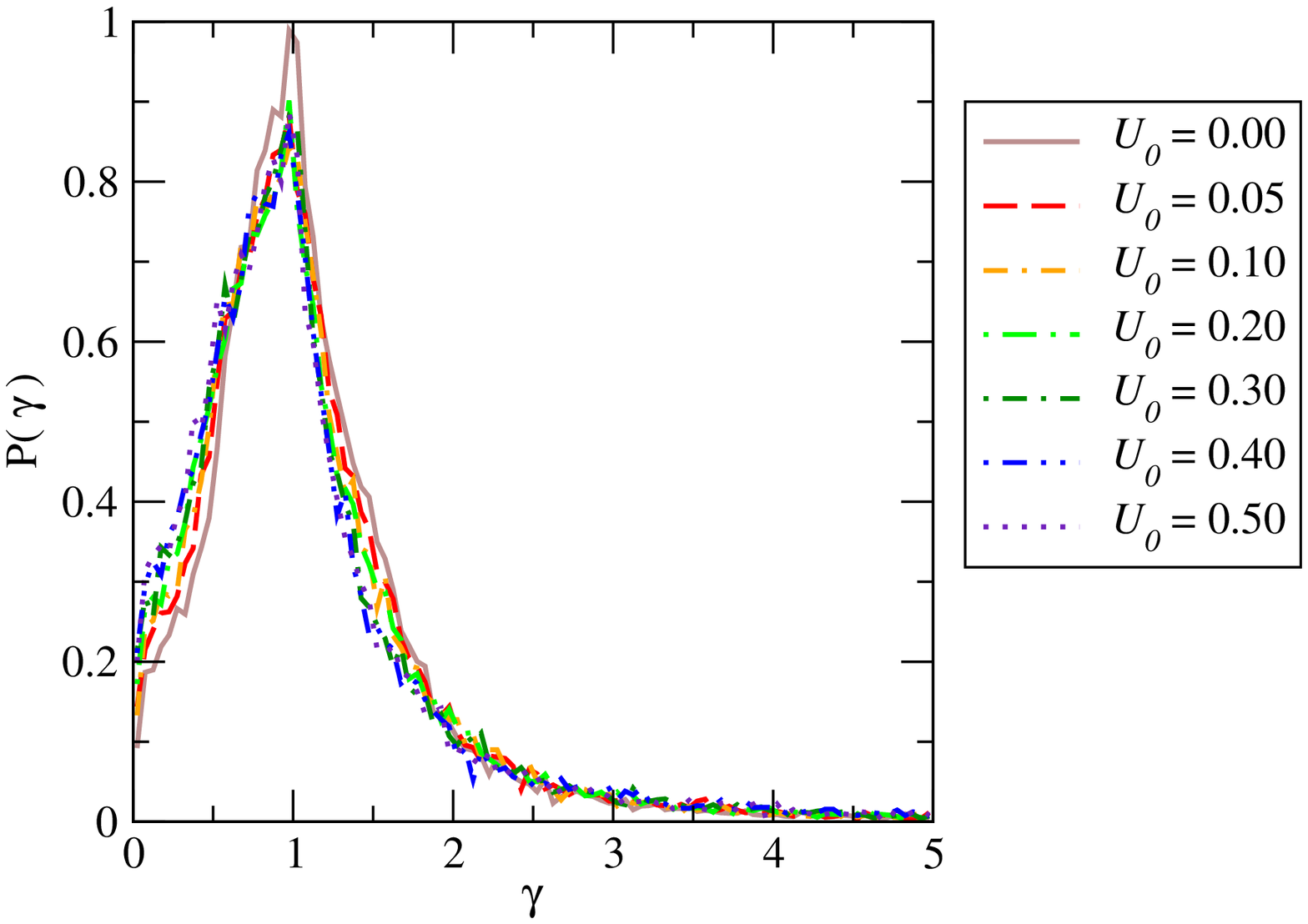}
	\label{constrained_gamma}
}
\caption{
	(Color)
	Distributions for the ratio of child-to-parent length $\gamma$.
	(a)~Distributions of $\gamma$ in mouse lung and human head and torso.
	(b)~Distributions of $\gamma$ for several thresholds of hierarchical unbalance $U_0$,  optimized solely for total network length $L$ ${[F(1, 0)]}$.
}
\label{gamma_distributions}
\end{figure*}

In Fig. \ref{gamma_dicom} we show the network-wide distribution for the ratio of child-to-parent lengths $\gamma = \ell_c/\ell_p$ for child $c$ with parent $p$.
Although there is the tendency that $\gamma < 1$, some child segments have a relatively shorter parent.
Although slight, an increased threshold for $U_0$ shifts more child segments to be shorter than their associated parent (see Fig. \ref{constrained_gamma}).
Independent of the threshold, the nonzero variance of this distribution shows that $\gamma$ is not constant throughout the network.

\begin{acknowledgments}
We would like to thank Kristina I. Boström, MD, PhD and  Yucheng Yao, MD, PhD for sharing their data of mouse lung vasculature.
We would also like to thank Daniel Ennis, PhD for sharing his data of human head and torso vasculature and Mitchell Johnson for his work in developing the software to analyze vascular images.
We are grateful to Eric Deeds, PhD and Tom Kolokotrones, MD, MPH, PhD for engaging in stimulating discussions about the work.
We would also like to thank Elif Tekin for her patient help in refining the presentation in this article.
\end{acknowledgments}

\bibliography{space-filling_trees_prx}

\end{document}